\documentclass[showpacs,preprintnumbers,amsmath,amssymb,aps]{revtex4}

\usepackage{graphicx}
\usepackage{dcolumn}
\usepackage{bm}

\begin{document}


\title{Theory of the Trojan-Horse Method}

\author{S. Typel}
\affiliation{
Department of Physics and Astronomy and
National Superconducting Cyclotron Laboratory,
Michigan State University, 
East Lansing, Michigan 48824-1321es
}

\author{G. Baur}
\affiliation{
Institut f\"{u}r Kernphysik,
Forschungszentrum J\"{u}lich,
52425 J\"{u}lich, Germany
}

\date{\today}

\begin{abstract}
The Trojan-Horse method is an indirect approach to determine
the energy dependence of S-factors of astrophysically relevant
two-body reactions. This is accomplished by studying  closely
related three-body reactions under quasi-free scattering conditions.
The basic theory of the Trojan-Horse method is developed starting from
a post-form distorted wave Born approximation of the T-matrix
element. In the surface approximation the cross section of the three-body reaction 
can be related to the S-matrix elements of the two-body reaction.
The essential feature of the Trojan-Horse method is the effective
suppression of the Coulomb barrier at low energies for the
astrophysical reaction leading to finite cross sections at the
threshold of the two-body reaction. In a modified
plane wave approximation the relation 
between the two-body and three-body cross sections becomes very transparent.
The appearing Trojan-Horse integrals are
studied in detail. 
\end{abstract}

\pacs{24.50.+g, 24.10.-i, 25.10.+s}
\maketitle

\mbox{} \hfill \dots $\kappa \epsilon \kappa \alpha \lambda \upsilon
\mu \mu \epsilon \nu o \iota$ $\iota \pi \pi \omega$. \cite{Homer}

\section{Introduction}

Nuclear reaction rates are an indispensable ingredient of many astrophysical models.
They have to be known with sufficient accuracy in order to
understand quantitatively the evolution of the universe, stars and other objects
in the cosmos \cite{Rol88,Thi01,Rol01}.
Ideally, reaction cross sections are directly 
measured in the laboratory. But, with few exceptions and despite many
experimental efforts, the relevant low
energy range cannot be reached in direct experiments \cite{LUN02,Bon99,Jun98}. 
Cross sections for reactions with charged particles
rapidly become very small with decreasing energy of the colliding
nuclei due to the repulsive Coulomb interaction. Extrapolations of the cross section
$\sigma(E)$ to low energies
from results at higher energies accessible to experiments are often needed.
This is accomplished with the help of the astrophysical S-factor
\begin{equation} \label{Sfac}
 S(E) = \sigma(E) \: E \: \exp(2\pi \eta)
\end{equation}
where $E$ is the c.m.\ energy and
$\eta = Z_{1}Z_{2}e^{2}/(\hbar v)$ is the Sommerfeld parameter
which depends on the charge numbers $Z_{1}$, $Z_{2}$ of the colliding nuclei
and their relative velocity $v$ in the entrance channel.
The S-factor shows a much weaker energy dependence than the cross section
$\sigma(E)$ because the main effect of the
penetrability through the  Coulomb barrier is compensated by
the increase of the exponential factor.  
The extrapolation process introduces uncertainties and important 
contributions to the cross sections, like resonances, 
can be missed. Additionally, direct laboratory 
experiments are affected by electron screening, which effectively
reduces the Coulomb barrier between the nuclei and enhances the measured
laboratory cross section \cite{Ass87,Sho93,Lio01}. A correction has to be applied
to obtain the cross section for bare nuclei. This effect does not seem to be 
completely understood yet and independent information on low energy
cross sections is valuable in order to develop a quantitative
description of electron screening. In astrophysical applications
one has, in addition, to account for the screening under
stellar conditions.

In recent years several indirect methods have been developed 
to extract cross sections
relevant to astrophysics  from other types of
experiments. In these alternative approaches
the astrophysical relevant two-body reaction is generally 
replaced by a suitably chosen
three-body reaction. The relation between the reactions is established with the
help of nuclear reaction theories. Without doubt, this process will 
introduce some uncertainties,
but valuable information on the astrophysical
reaction can be obtained. Also, the errors are independent 
from that of the direct approach.
Of course, the indirect methods have to be validated by studying well known
reactions before firm conclusions can be drawn from indirect experiments
in cases where direct measurements are not feasible; see also
the minireview \cite{Aus02}.

The Coulomb dissociation method has become a valuable tool for
extracting low-energy cross sections of radiative capture reactions 
$a(b,\gamma)c$
by studying the inverse process of photo dissociation
$c(\gamma,b)a$ \cite{Bau94}. Instead of using
real photons, the Coulomb field of a highly charged target $X$ acts as
a source of virtual photons during a scattering process which leads to
the breakup reaction $X(c,ab)X$ with three particles in the final state. 
Due to the high flux of virtual photons, cross sections at the small
relevant energies in the two-body system are strongly enhanced as compared
to the capture reaction. 
Another approach is the ANC method that tries to extract the 
asymptotic normalization coefficient of a nuclear ground state wave function
by studying various combinations of transfer reactions involving
this nucleus at low energies \cite{Azh01,Xu94}.
The coefficient can be used to determine theoretically the astrophysical S-factor 
for radiative capture reactions at zero energy. However, these indirect
approaches are limited to astrophysically relevant reactions 
where a photon is involved.

The observation of a similarity between cross sections for two-body and
closely related three-body reactions under certain kinematical conditions
\cite{Fuc71}
led to the introduction of the Trojan-Horse 
method (THM) \cite{Bau86}, see also \cite{Bau84,Bau76}.
The aim of the THM is to extract the cross section
of an astrophysically relevant two-body reaction
\begin{equation} \label{APreac}
 A + x \to C + c
\end{equation}
from a suitably chosen reaction
\begin{equation} \label{THreac}
 A + a \to C + c + b
\end{equation}
with three particles in the final state assuming that the Trojan Horse
$a$ is composed predominantly of clusters $x$ and $b$. 
The kinematical conditions are chosen such that the momentum transfer
to the nucleus $b$ is small during the reaction. Therefore $b$ can be considered as
a spectator to the reaction of $A$ and $x$. 
This process is often referred to as a quasi-free scattering.
In a selected part of the available three-body phase space it
is known to dominate over other reaction mechanism like sequential breakup processes.
In the past quasi-free scattering  
has been used to extract information on momentum distributions of the
nucleus $a$, i.e. the ground state wave function in momentum space,
employing a plane-wave impulse approximation (PWIA) in the
theoretical description \cite{Jai70}. In this approach the cross section
of reaction (\ref{THreac})
factorizes into a kinematical factor, the ground state momentum distribution of
nucleus $a$ and an off-shell cross section of
reaction (\ref{APreac}) that is assumed to be known. 
On the other hand 
the two-body cross section can be extracted from the cross section
of reaction (\ref{THreac}) if the momentum distribution of the Trojan Horse
$a$ is known with 
sufficient accuracy and a relation between the off-shell and on-shell two-body
cross sections can be established. The selection of different spectators
$b$ and thus Trojan Horses
$a$ allows additional checks of the underlying assumptions of the method.

Unlike the Coulomb dissociation method and the ANC method which 
are limited to radiative processes, 
the THM can be applied to reactions where no photon is involved.
The essential feature of the THM is the actual suppression of the Coulomb
barrier in the cross section of the
two-body reaction. The cross section of the three-body reaction
is not reduced when the c.m.\ energy in the $A+x$ system approaches zero as in 
reaction (\ref{APreac}). The energy in the entrance channel of (\ref{THreac})
can be around or above the Coulomb barrier and effects from electron screening 
are negligible. Nevertheless, very small energies in the  reaction (\ref{APreac})
can be reached.

The feasibility of the THM was studied in several experiments involving
various reactions during the last years
\cite{Mus01,Spi01,Che96,Lat01,Ali00,Spi99,Cal97,Spi00,Pel00}. 
In earlier evaluations of the experiments
some simplifying assumptions were made in the theoretical description.
The off-shell two-body cross section as appearing in the PWIA was considered
as the bare nuclear cross section. It was converted to the on-shell
two-body cross section by correcting for the Coulomb penetration
in a heuristic approximation.
Basic considerations for the theoretical description of the THM 
were developed in \cite{Typ00} but 
the deduced relation between the cross sections for reactions
(\ref{APreac}) and (\ref{THreac}) were not directly applicable to the experiments.

In this paper the theory of the THM is developed in certain approximations
that allow to establish a simple connection between the cross sections
of the three-body and two-body reactions.
In Section \ref{RT} the reaction theory is formulated and the relation
of the T-matrix element of the three-body reaction 
with the S-matrix elements of the two-body reaction is found.
In connection with a plane wave approximation, fundamental
TH integrals appear that are discussed in Section \ref{THI}.
In the following two sections
expressions for scattering amplitudes and cross sections
are derived for spinless particles and
then for the general case of particles with spin. 
The TH Coulomb scattering amplitude that is relevant
in the indirect study of elastic scattering processes is treated
in Section \ref{SfTHC}. Kinematical conditions, 
the energy dependence of cross sections
and applications ot the THM are discussed in Section \ref{Sappl}.
A summary and an outlook are presented in the last Section.
Details of the analytical calculations are given in the Appendices.
The evaluation of actual Trojan-Horse experiments with
the present theory is beyond the scope of this paper und 
is subject to a detailed treatment in separate studies.

\section{Reaction Theory}
\label{RT}

\begin{figure}
\includegraphics[width=80mm]{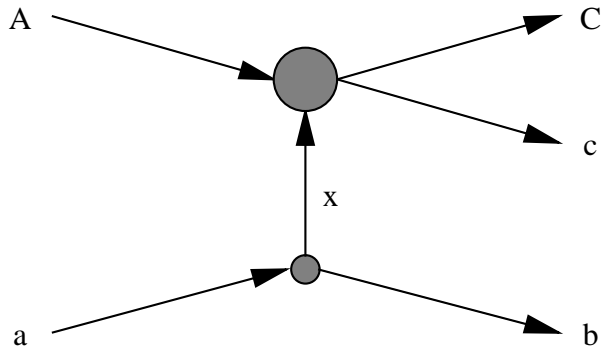}
\caption{\label{fig00} Transfer of particle $x$ in the Trojan-Horse
reaction (\ref{THreac}).}
\end{figure}

In the three-body reaction (\ref{THreac}) the particle $x$ is transferred
from the Trojan Horse $a$ to the nucleus $A$ leading to the 
$C+c$ final state with the spectator $b$ of the Trojan Horse remaining,
see Fig.\ \ref{fig00}.
This reaction can be considered as a transfer to the continuum, where
the inelastic reaction $A+x \to C+c$ can happen in addition
to the elastic $A+x$ scattering.
It is customary to describe such a direct reaction, i.e.\
$A+a \to C+c+b$, with the help of
a distorted wave Born approximation. It is not required in the theoretical 
approach that $A+x \neq C+c$; elastic scattering processes in the two-body
system can be treated in the formalism, too.
Effects from the antisymmetrization of the wave functions will be neglected
in the present treatment; they are expected to be small.

\subsection{Coordinate Systems and Cross Sections}

\begin{figure}
\includegraphics[width=180mm]{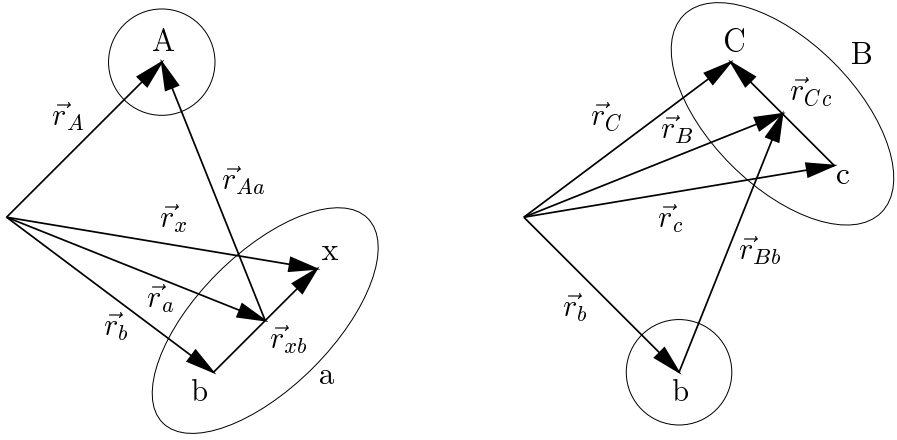}
\caption{\label{fig01} Coordinate vectors in the initial and final partitions
of the Trojan-Horse reaction (\ref{THreac}).}
\end{figure}

In a three-body system various sets of Jacobi coordinates are
used to specify the positions of the particles. In the theoretical
description we encounter the sets
\begin{equation}
 \vec{r}_{xb} = \vec{r}_{x} - \vec{r}_{b} \: ,
 \qquad
 \vec{r}_{Aa} = \vec{r}_{A} - \vec{r}_{a}
 = \vec{r}_{A} - \frac{m_{x}\vec{r}_{x}+m_{b}\vec{r}_{b}}{m_{x}+m_{b}}
\end{equation}
in the initial partition
and
\begin{equation}
 \vec{r}_{Cc} = \vec{r}_{C} - \vec{r}_{c} \: ,
 \qquad
 \vec{r}_{Bb} = \vec{r}_{B} - \vec{r}_{b}
 = \frac{m_{C}\vec{r}_{C}+m_{c}\vec{r}_{c}}{m_{C}+m_{c}} - \vec{r}_{b}
\end{equation}
in the final partition. The symbol $B$ denotes the $C+c=A+x$ system.
The coordinate vectors are shown in Fig.\ \ref{fig01}.
The corresponding relative momenta and wave vectors are given by
\begin{equation}
 \vec{p}_{ij} = \hbar \vec{k}_{ij} = 
 \frac{m_{j}\vec{p}_{i}-m_{i}\vec{p}_{j}}{m_{i}+m_{j}}
\end{equation}
for nuclei $i$ and $j$ with masses $m_{i},m_{j}$ and momenta
$\vec{p}_{i}, \vec{p}_{j}$ in the laboratory system, respectively.
With the help of the kinetic energies
\begin{equation}
 E_{ij} = \frac{p_{ij}^{2}}{2\mu_{ij}} 
\end{equation} 
where the reduced masses 
\begin{equation}
 \mu_{ij} = \frac{m_{i}m_{j}}{m_{i}+m_{j}}
\end{equation}
appear, energy conservation in the two-body reaction (\ref{APreac})
can be expressed as
\begin{equation} \label{ec2}
 E_{Ax}  =  E_{Cc} - Q_{2}
\end{equation}
with the Q-value
\begin{equation}
 Q_{2} = (m_{A} + m_{x} - m_{C} - m_{c})c^{2}
\end{equation}
and similarly
\begin{equation} \label{ec3}
  E_{Aa}  =  E_{Cc} + E_{Bb} - Q_{3}
 = E_{Ax} + E_{Bb} + Q_{2} - Q_{3}
\end{equation}
with
\begin{equation}
 Q_{3} = (m_{A} + m_{a} - m_{C} - m_{c} - m_{b})c^{2}
\end{equation}
in case of the three-body reaction. 

The general cross section for reaction (\ref{THreac}) with three particles in
the final state depends on the choice of the independent variables
and the reference system. In the c.m.\ system the
energy $E_{Cc}$ and the directions of the Jacobi momenta 
$\vec{p}_{Bb}$ and $\vec{p}_{Cc}$ completely specify the kinematical
conditions for a given projectile energy. Then the differential cross
section takes the form
\begin{equation} \label{d3scm}
 \frac{d^{3}\sigma}{dE_{Cc}d\Omega_{Cc}d\Omega_{Bb}}
 =  K_{\rm cm} \left|  T_{fi} (\vec{k}_{Cc},\vec{k}_{Bb};\vec{k}_{Aa})
 \right|^{2} 
\end{equation}
with the c.m. kinematical factor
\begin{equation}
 K_{\rm cm} = \frac{\mu_{Aa}\mu_{Bb}\mu_{Cc}}{(2\pi)^{5}\hbar^{7}}
 \frac{p_{Bb}p_{Cc}}{p_{Aa}} 
\end{equation}
and the T-matrix element $T_{fi}$.
In an actual experiment, the particles $C$ and $c$ are usually detected 
in the laboratory system
and it is more convenient to use
\begin{equation} \label{d3slab}
 \frac{d^{3}\sigma}{dE_{C}d\Omega_{C}d\Omega_{c}}
 =  K_{\rm lab} \left|  T_{fi} (\vec{k}_{Cc},\vec{k}_{Bb};\vec{k}_{Aa})
 \right|^{2}
\end{equation}
with the kinematical factor
\begin{equation}
 K_{\rm lab} = \frac{\mu_{Aa}m_{C}}{(2\pi)^{5}\hbar^{7}}
 \frac{p_{C}p_{c}^{2}}{p_{Aa}}
 \left[ \left( \frac{\vec{p}_{Bb}}{\mu_{Bb}} 
 - \frac{\vec{p}_{Cc}}{m_{c}} \right) \cdot \frac{\vec{p}_{c}}{p_{c}}\right]^{-1}
\: .
\end{equation}
The laboratory differential cross section depends on the scattering angles of the
nuclei $C$ and $c$ and the energy $E_{C}$. 
Again, these quantities, together
with the beam energy, specify the kinematical conditions completely.
The nuclei $A$ and $a$ can be projectile and target or vice versa.
In case of particles with spin appropriate averages over initial
states and sums over final states have to be considered.

\subsection{Approximations of the T-Matrix Element}

The T-matrix element $T_{fi}$ in (\ref{d3scm}) and (\ref{d3slab})
contains all the essential information relevant to the scattering process.
It has to be calculated in a suitable approximation that allows to
find the connection to the cross section of the astrophysical 
relevant reaction (\ref{APreac}).

The Trojan-Horse reaction (\ref{THreac}) has the
form of a usual two-body reaction
\begin{equation}
 A + a \to B + b
\end{equation}
if the $C+c$ system is considered as an excited state of the
compound system $B$.
The exact T-matrix element for  this reaction 
is given in the post-form description by
\begin{equation} \label{texact}
 T_{fi} 
 = \langle \exp\left(i\vec{k}_{Bb}\cdot \vec{r}_{Bb}\right) 
  \phi_{B} \phi_{b}  | V_{Bb} | \Psi^{(+)}_{Aa}(\vec{k}_{Aa},\vec{r}_{Aa}) \rangle
\end{equation}
with the exact scatttering wave function $\Psi^{(+)}_{Aa}(\vec{k}_{Aa},\vec{r}_{Aa})$ 
in the initial state and the potential $V_{Bb}$ between $B$ and $b$ in the
final state \cite{New66}. The wavefunction $\phi_{i}$ of a nucleus $i$ depends only
on internal coordinates which are not given explicitly. The relative motion
of $B$ and $b$ is described by a plane wave with momentum $\hbar \vec{k}_{Bb}$.
Applying the Gell-Mann Goldberger transformation \cite{Gel53},
the T-matrix element (\ref{texact}) assumes the form
\begin{equation} \label{texact2}
 T_{fi} 
 = \langle \chi^{(-)}_{Bb}(\vec{k}_{Bb}, \vec{r}_{Bb}) 
  \phi_{B} \phi_{b}  | V_{Bb} - U_{Bb} | \Psi^{(+)}_{Aa}(\vec{k}_{Aa}, \vec{r}_{Aa}) 
  \rangle
\end{equation}
which is still an exact relation. For a detailed derivation
see \cite{Typ00}.
Here, $\chi^{(-)}_{Bb}(\vec{k}_{Bb}, \vec{r}_{Bb})$ 
is a distorted wave for the $B-b$ relative
motion generated by a suitably chosen optical potential $U_{Bb}$ that only depends on
$\vec{r}_{Bb}$ but not on internal coordinates in contrast to $V_{Bb}$.
In the THM the wave function $\phi_{B}$ does not describe  a bound state but a
complete scattering state 
\begin{equation}
 \phi_{B} =  \Psi^{(-)}_{Cc}(\vec{k}_{Cc},\vec{r}_{Cc})
\end{equation}
since the system $B=C+c$ is in the continuum.

In the post-form distorted-wave Born approximation (DWBA) the T-matrix element
(\ref{texact2})
is replaced by
\begin{equation} \label{TDWBA}
 T_{fi}^{DWBA} (\vec{k}_{Cc},\vec{k}_{Bb};\vec{k}_{Aa})
 = \langle \chi^{(-)}_{Bb}(\vec{k}_{Bb}, \vec{r}_{Bb}) 
  \Psi^{(-)}_{Cc}(\vec{k}_{Cc}, \vec{r}_{Cc}) \phi_{b} 
 | V_{xb} |
 \chi^{(+)}_{Aa}(\vec{k}_{Aa}, \vec{r}_{Aa}) \phi_{A} \phi_{a} \rangle
\end{equation}
where the exact wave function $\Psi^{(+)}_{Aa}(\vec{k}_{Aa},\vec{r}_{Aa})$
is replaced by a distorted wave $\chi^{(+)}_{Aa}(\vec{k}_{Aa},\vec{r}_{Aa})$ 
for the $A-a$ relative motion. Additionally, the potential $V_{Bb}-U_{Bb}$ is
approximated by $V_{xb}$ \cite{Typ00,Aus70}.
This is the usual starting point of T-matrix calculations for direct 
transfer reactions.

In the next step, the so-called surface approximation is applied
which is essential to the THM.
For small distances between the colliding nuclei the optical potentials
are usually strongly absorptive and only reactions at the surface of the nuclei
contribute significantly to the matrix element. Therefore the full scattering
wave function $\Psi^{(-)}_{Cc}(\vec{k}_{Cc},\vec{r}_{Cc})$ can be replaced by its
asymptotic form for radii larger than a suitably chosen cutoff radius $R$
that is larger than the range of the nuclear
potential.
It is typically in the range of the sum of the
radii of the two colliding nuclei. 
The interior part of $ \Psi^{(-)}_{Cc}(\vec{k}_{Cc},\vec{r}_{Cc})$ is set to zero.
The validity of the surface approximation was checked
in \cite{Kas82};
it was found to be quite good for the (d,p) reaction at $E_{d}=26$~MeV.
The asymptotic form of the scattering wave function
for $r_{\alpha} > R$, where $\alpha=Ax, Cc, \dots$
is a possible partition of the system $B$, is given by
\begin{equation} \label{wfasym}
 \Psi^{(\pm)}_{Cc,{\rm asym}}(\vec{k}_{Cc},\vec{r}_{Cc})
 = \frac{4\pi}{k_{Cc}} 
 \sum_{\alpha} \sum_{lm} \frac{\xi_{l}^{(\pm)}(\alpha)}{r_{\alpha}}
 i^{l} Y_{lm}(\hat{r}_{\alpha}) Y_{lm}^{\ast}(\hat{k}_{Cc}) \phi_{\alpha}
\end{equation}
with radial wave fuctions
\begin{eqnarray}
 \xi_{l}^{(+)} (\alpha) & = &
 \frac{1}{2i} \sqrt{\frac{v_{Cc}}{v_{\alpha}}}
 \left[ S^{l}_{\alpha Cc} 
 u_{l}^{(+)}(\eta_{\alpha};k_{\alpha}r_{\alpha})
 - \delta_{\alpha Cc} u_{l}^{(-)}(\eta_{\alpha};k_{\alpha}r_{\alpha})
 \right] \: ,
 \\ 
 \xi_{l}^{(-)} (\alpha) & = &
\xi_{l}^{(+)\ast} (\alpha)
\end{eqnarray}
where  the  Coulomb wave functions \cite{Abr70}
\begin{eqnarray} \label{Cwf}
  u_{l}^{(\pm)}(\eta;kr)
 & = & e^{\mp i \sigma_{l}(\eta)}
 \left[G_{l}(\eta;kr) \pm i 
 F_{l}(\eta;kr) \right]
 \\ \nonumber & \to & 
 \exp \left[\pm i \left( kr 
 - \eta \ln (2 kr) - \frac{l\pi}{2}\right) \right]
\end{eqnarray}
appear. The Sommerfeld parameter
\begin{equation}
 \eta_{ij} = \frac{Z_{i}Z_{j}e^{2}}{\hbar v_{ij}}
\end{equation}
depends on the charge numbers $Z_{i}, Z_{j}$ of
nuclei $i,j$
and their relative velocity $v_{ij}= p_{ij}/\mu_{ij}$
in the partition $\alpha = ij$.
The Coulomb phase shifts are given by $\sigma_{l}(\eta) = \arg \Gamma(1+l+i\eta)$.
The S-matrix elements $S^{l}_{\alpha Cc}$ in the radial wave functions
completely describe the two-body scattering process. They are on-shell
quantities and the momenta $k_{\alpha}$ are derived from energy-conservation
in the two-body reaction. E.g., in the $\alpha=Ax$ partition we have
$k_{Ax}= \sqrt{2\mu_{Ax}E_{Ax}}/\hbar$ with the energy $E_{Ax}$ from
relation (\ref{ec2}). 

The potential $V_{xb}$ appearing in the T-matrix element (\ref{TDWBA})
describes the
interaction between $x$ and $b$ in the Trojan Horse $a$. Assuming a simple
cluster picture of $a$ and neglecting contributions from
excited states of $x$ and $b$ or
other partitions in the ground state, 
the momentum amplitude $W(\vec{q})$ of the product
\begin{equation}
 V_{xb}(\vec{r}_{xb}) \phi_{a} (\vec{r}_{xb})
 = \int \frac{d^{3}q}{(2\pi)^{3}} \: W(\vec{q})
 \exp \left( i \vec{q} \cdot \vec{r}_{xb} \right) \phi_{x} \phi_{b}
\end{equation}
can be introduced and
the T-matrix element assumes the form
\begin{eqnarray} \label{TWME}
 \lefteqn{T_{fi}^{TH}(\vec{k}_{Cc},\vec{k}_{Bb};\vec{k}_{Aa}) =
 \int \frac{d^{3}q}{(2\pi)^{3}} \: W(\vec{q})}
 \\ \nonumber & & 
  \langle \chi^{(-)}_{Bb}(\vec{k}_{Bb}, \vec{r}_{Bb}) 
  \Psi^{(-)}_{Cc,{\rm asym}}(\vec{k}_{Cc}, \vec{r}_{Cc}) 
 | \exp \left( i \vec{q} \cdot \vec{r}_{xb} \right) 
 \chi^{(+)}_{Aa}(\vec{k}_{Aa}, \vec{r}_{Aa}) \phi_{A} \phi_{x} \rangle
 \: .
\end{eqnarray}
The assumption $W\equiv$ const.\ corresponds to the zero-range approximation which
is frequently employed in the calculation of DWBA T-matrix elements.
In this case, the integration over $\vec{q}$ leads to a $\delta$-function
in the variable $\vec{r}_{xb}$ and
the actual dimension of the appearing integral is reduced from six to three.
However, in general the full momentum dependence of the amplitude $W$ 
has to be considered. 
With the help of the Schr\"{o}dinger equation 
the momentum amplitude $W$ is related by
\begin{equation}
 W(\vec{q}) = - \left( \varepsilon_{a} 
 + \frac{\hbar^{2} q^{2}}{2\mu_{xb}}\right) 
\Phi_{a}(\vec{q})
\end{equation}
to the ground state momentum wave function
\begin{equation}
 \Phi_{a}(\vec{q}) = \langle  
 \exp \left( i \vec{q} \cdot \vec{r}_{xb} \right) \phi_{x} \phi_{b}
 |   \phi_{a} (\vec{r}_{xb}) \rangle
\end{equation}
of the nucleus $a$. The energy $\varepsilon_{a} = (m_{x}+m_{b}-m_{a})c^{2}
 = Q_{2}-Q_{3} >0$ is
the binding energy of $a$ with respect to the $x+b$ threshold.

The integration over the internal coordinates in the matrix element
(\ref{TWME}) selects the $\alpha = Ax$
partition of the asymptotic wave function (\ref{wfasym}) and
the T-matrix element in the TH approximation can be written as
\begin{eqnarray} \label{TTHgen}
  \lefteqn{T_{fi}^{TH}(\vec{k}_{Cc},\vec{k}_{Bb};\vec{k}_{Aa}) =
  \frac{1}{2ik_{Cc}} \sqrt{\frac{v_{Cc}}{v_{Ax}}} \sum_{l} (2l+1)}
 \\ \nonumber & & 
 \left[ S^{l}_{Ax Cc} \:  U_{l}^{(+)} (\vec{k}_{Bb}, \vec{k}_{Cc}, \vec{k}_{Aa})
 - \delta_{Ax Cc} \: U_{l}^{(-)}(\vec{k}_{Bb}, \vec{k}_{Cc}, \vec{k}_{Aa})
 \right]  \: .
\end{eqnarray}
It resembles a scattering amplitude of the reaction $C+c \to A+x$ except
for the functions
\begin{eqnarray} \label{Udef}
 \lefteqn{U_{l}^{(\pm)}(\vec{k}_{Bb}, \vec{k}_{Cc}, \vec{k}_{Aa})
 =  \frac{4\pi (-i)^{l}}{(2l+1)} \sum_{m}  Y_{lm}(\hat{k}_{Cc}) 
 \int \frac{d^{3}q}{(2\pi)^{3}} \: W(\vec{q})}
 \\ \nonumber & & 
  \langle \chi^{(-)}_{Bb}(\vec{k}_{Bb}, \vec{r}_{Bb}) 
  \frac{\theta(r_{Ax}-R)}{r_{Ax}} 
 u_{l}^{(\mp)}(\eta_{Ax};k_{Ax}r_{Ax}) Y_{lm}(\hat{r}_{Ax})
 |  \exp \left( i \vec{q} \cdot \vec{r}_{xb} \right) 
 \chi^{(+)}_{Aa}(\vec{k}_{Aa}, \vec{r}_{Aa}) \rangle 
\end{eqnarray}
that describe the angular distribution and the
momentum dependence
due to the presence of the
spectator $b$ in the reaction (\ref{THreac}).
The unit step function $\theta$ accounts for the surface 
approximation by eliminating the contributions at small radii.
With equation (\ref{TTHgen}) the relation between
the cross section (\ref{d3slab}) of the three-body reaction (\ref{THreac}) 
and the S-matrix elements of the two-body reaction 
is directly established. It is possible to extract the energy dependence
of the S-matrix elements from experimental three-body cross sections, 
at least in principle. For an
actual application of the TH method, it is convenient to introduce
additional approximations that lead to a formulation with
a direct relation between the cross sections of the two-body
and three-body reactions similar to the PWIA.

Even with the surface approximation, the remaining matrix element
involves a six-dimensional integration in the Jacobi-coordinates
which is an extensive computational task. 
Various approximation can be introduced to simplify the calculation.
It is useful to expand 
the distorted wave in the initial state in a Taylor series
\begin{eqnarray} \label{tayl}
\chi^{(+)}_{Aa}(\vec{k}_{Aa}, \vec{r}_{Aa}) & = & 
 \left( 1 + \left( \vec{r}_{Aa}-\vec{r}_{Ax}\right) \cdot \vec{\nabla}
 + \dots \right) 
 \chi^{(+)}_{Aa}(\vec{k}_{Aa}, \vec{r}_{Ax})
 \\ \nonumber & \approx &
 \exp \left[ i \left( \vec{r}_{Aa}-\vec{r}_{Ax}\right)\cdot \vec{K}\right]
 \chi^{(+)}_{Aa}(\vec{k}_{Aa}, \vec{r}_{Ax})
\end{eqnarray}
where the wave vector $\vec{K}$ replaces the derivative 
$-i\vec{\nabla}$ with respect to the spatial coordinates.
In the so called local momentum approximation \cite{Bra74,Shy85} the modulus
of $\vec{K}$ is determined by the actual kinetic energy of the $A-a$ relative
motion at a certain distance $R_{Aa}$. The direction of $\vec{K}$ is assumed
to be same as the asymptotic momentum $\vec{k}_{Aa}$.
If $\chi^{(+)}_{Aa}$ is a plane wave and 
$\vec{K}=\vec{k}_{Aa}$ the relation (\ref{tayl}) is obviously exact.
Considering the relations
\begin{equation}
 \vec{r}_{xb}  =  \vec{r}_{Bb} - \alpha  \vec{r}_{Ax} \: ,
 \qquad
 \vec{r}_{Aa}  =  \beta  \vec{r}_{Bb}  + (1-\alpha \beta) \vec{r}_{Ax}
\end{equation}
of the Jacobi coordinates with factors
\begin{equation}
 \alpha = \frac{m_{A}}{m_{A}+m_{x}} \: ,
 \qquad 
 \beta = \frac{m_{b}}{m_{b}+m_{x}} \: ,
\end{equation}
the matrix element in equation (\ref{TWME})
factorizes and the T-matrix element in the TH approximation
assumes the form
\begin{equation} \label{TWME2}
 T_{fi}^{TH}(\vec{k}_{Cc},\vec{k}_{Bb};\vec{k}_{Aa}) =
 \int \frac{d^{3}q}{(2\pi)^{3}} \: {\cal W}(\vec{q},\vec{k}_{Bb})
 \: {\cal M}(\vec{q}, \vec{k}_{Cc}, \vec{k}_{Aa})
\end{equation}
with the generalized momentum amplitude
\begin{equation} \label{MMD}
 {\cal W}(\vec{q},\vec{k}_{Bb}) =  W(\vec{q}) \: 
  \langle \chi^{(-)}_{Bb}(\vec{k}_{Bb}, \vec{r}_{Bb}) 
 | \exp \left( i \vec{Q} \cdot \vec{r}_{Bb} \right) \rangle 
\end{equation}
and the matrix element
\begin{eqnarray} \label{defM0}
 {\cal M}(\vec{q}, \vec{k}_{Cc}, \vec{k}_{Aa}) =
 \langle  \Psi^{(-)}_{Cc,{\rm asym}}(\vec{k}_{Cc}, \vec{r}_{Cc}) 
 | \exp \left( - i \alpha \vec{Q}  \cdot \vec{r}_{Ax} \right)
 \chi^{(+)}_{Aa}(\vec{k}_{Aa}, \vec{r}_{Ax}) \phi_{A} \phi_{x} \rangle
\end{eqnarray}
with $\vec{Q} = \vec{q} + \beta \vec{K}$.
The integration over the internal coordinates leads to the expression
\begin{eqnarray} \label{defM}
 \lefteqn{{\cal M}(\vec{q}, \vec{k}_{Cc}, \vec{k}_{Aa}) =
\frac{1}{2ik_{Cc}} \sqrt{\frac{v_{Cc}}{v_{Ax}}} \sum_{l} (2l+1)}
 \\ \nonumber & & 
 \left[ S^{l}_{Ax Cc} \:  {\cal U}_{l}^{(+)} (\vec{q}, \vec{k}_{Cc}, \vec{k}_{Aa})
 - \delta_{Ax Cc} \: {\cal U}_{l}^{(-)}(\vec{q}, \vec{k}_{Cc}, \vec{k}_{Aa})
 \right]  
\end{eqnarray}
with the integrals
\begin{eqnarray} \label{calUdef}
{\cal U}_{l}^{(\pm)} (\vec{q}, \vec{k}_{Cc}, \vec{k}_{Aa})& = & 
 \frac{4\pi (-i)^{l}}{(2l+1)}
 \int d^{3}r_{Ax} \: \frac{\theta(r_{Ax}-R)}{r_{Ax}} 
 u_{l}^{(\pm)}(\eta_{Ax};k_{Ax}r_{Ax})
 \\ \nonumber & & 
 \exp \left( - i \alpha \vec{Q}  \cdot \vec{r}_{Ax} \right)
 \chi^{(+)}_{Aa}(\vec{k}_{Aa}, \vec{r}_{Ax})
 \sum_{m}  Y_{lm}^{\ast} (\hat{r}_{Ax}) Y_{lm}(\hat{k}_{Cc}) \: .
\end{eqnarray}
The functions $U_{l}^{(\pm)}$ in the TH T-matrix element (\ref{TTHgen}) 
assume the form
\begin{equation} 
 U_{l}^{(\pm)}(\vec{k}_{Bb}, \vec{k}_{Cc}, \vec{k}_{Aa})
 = \int \frac{d^{3}q}{(2\pi)^{3}} \: {\cal W}(\vec{q},\vec{k}_{Bb})
 \: {\cal U}_{l}^{(\pm)} (\vec{q}, \vec{k}_{Cc}, \vec{k}_{Aa}) \: .
\end{equation}
The dependence
of $U_{l}^{(\pm)}$ on $k_{Cc}$ or equivalently $k_{Ax}$ together with
the energy dependence of the two-body S-matrix elements $S^{l}_{Ax Cc}$
leads to a finite cross section of the three-body reaction (\ref{THreac})
even when the threshold of the 
two-body reaction (\ref{APreac}) is reached. 
This is the essential feature of the TH method.
However this is not readily seen from the general 
expression (\ref{Udef}). A simplified formulation allows
to study the energy dependence more explicitly.

\subsection{Plane Wave Approximations}

If $\chi^{(-)}_{Bb}$ is assumed to be a plane wave, the matrix element
in the generalized momentum distribution (\ref{MMD}) leads to a $\delta$-function
in the variable $\vec{q}$
\begin{equation}
 {\cal W}(\vec{q},\vec{k}_{Bb}) =  W(\vec{q}) \: (2\pi)^{3} 
 \: \delta\left( \vec{q} - \vec{Q}_{Bb}\right)
\end{equation}
with 
\begin{equation}
 \vec{Q}_{Bb}  =  \vec{k}_{Bb} - \beta \vec{K} \: . 
\end{equation}
Then, the $\vec{q}$-integration is trivial;  the T-matrix element (\ref{TWME2})
factorizes explicitly into a momentum amplitude and
the matrix element (\ref{defM}) where the integrals (\ref{calUdef}) are
evaluated for $\vec{q} = \vec{Q}_{Bb}$.
This plane-wave approximation 
for $\chi^{(-)}_{Bb}$ is justified when the energy of the 
$Bb$ relative motion is large. 
When the spectator $b$ is a neutron, it is also often possible
to replace  $\chi^{(-)}_{Bb}$ by a plane wave.
On the other hand, when 
$\chi^{(-)}_{Bb}$ is replaced by a 
pure Coulomb scattering wave, e.g. if the $Bb$ relative energy is small,
the matrix element in (\ref{MMD})
can be calculated explicitly 
(see Appendix \ref{AppB}). However, in applications of the THM the c.m.\ energy 
in the $Bb$ system is generally large 
because of the large projectile energy and Q-value of the reaction
and the latter case is not relevant.

Independently from the treatment of the wave function $\chi^{(-)}_{Bb}$, a plane
wave approximation can be introduced for $\chi^{(+)}_{Aa}$
in the matrix element (\ref{defM0}).
In this case one finds 
\begin{equation} 
 {\cal M}(\vec{q}, \vec{k}_{Cc}, \vec{k}_{Aa}) =
 \langle
  \Psi^{(-)}_{Cc,{\rm asym}}(\vec{k}_{Cc}, \vec{r}_{Cc}) 
 | \exp \left( i \vec{Q}_{Aa}  \cdot \vec{r}_{Ax} \right)
   \phi_{A} \phi_{x} \rangle
\end{equation}
with  
\begin{equation}
 \vec{Q}_{Aa} =  \vec{k}_{Aa} -\alpha \beta \vec{K} - \alpha \vec{q} \: .
\end{equation}
The quantities (\ref{calUdef}) reduce to
\begin{eqnarray}  \label{calU2}
{\cal U}_{l}^{(\pm)} (\vec{q}, \vec{k}_{Cc}, \vec{k}_{Aa}) & = &
 \frac{4 \pi}{k_{Ax}Q_{Aa}} \:  P_{l}(\hat{Q}_{Aa}\cdot \hat{k}_{Cc})
 \: J_{l}^{(\pm)}(R,\eta_{Ax},k_{Ax},Q_{Aa})
\end{eqnarray}
with Legendre polynomials $P_{l}$ and the dimensionless
Trojan-Horse integrals
\begin{eqnarray} \label{defTHI}
 J_{l}^{(\pm)}(R,\eta_{Ax},k_{Ax},Q_{Aa})
 & = & k_{Ax} Q_{Aa}
  \int\limits_{R}^{\infty} dr_{Ax} \:  u_{l}^{(\pm)}(\eta_{Ax};k_{Ax}r_{Ax})
 \: r_{Ax} \:  j_{l}(Q_{Aa} r_{Ax}) \: .
\end{eqnarray}
The Coulomb wave functions $u_{l}^{(\pm)}$
as defined in (\ref{Cwf}) and
the regular spherical Bessel function $j_{l}$ of order $l$
appear in the radial integral.
It is convenient to introduce the decomposition
\begin{eqnarray} \label{THId}
 J_{l}^{(\pm)}(R,\eta_{Ax},k_{Ax},Q_{Aa})
 & = & e^{\mp i \sigma_{l}(\eta_{Ax})}
 \left [ J_{l}^{(G)}(R,\eta_{Ax},k_{Ax},Q_{Aa}) 
  \pm i  J_{l}^{(F)}(R,\eta_{Ax},k_{Ax},Q_{Aa})
 \right]
\end{eqnarray}
of the TH integrals into contributions with the regular and irregular
Coulomb wave functions. The integrals
are  discussed in detail in Section \ref{THI}.

In the following only the plane-wave approximation
for both $\chi^{(+)}_{Aa}$ and $\chi^{(-)}_{Bb}$ will be considered.
In this case the T-matrix element has the simple product form
\begin{eqnarray} \label{TTH}
 T_{fi}^{TH}(\vec{k}_{Cc},\vec{k}_{Bb};\vec{k}_{Aa}) & = & 
  W(\vec{Q}_{Bb})
 \langle   \Psi^{(-)}_{Cc,{\rm asym}}(\vec{k}_{Cc},\vec{r}_{Cc})  | 
 \exp \left( i \vec{Q}_{Aa} \cdot \vec{r}_{Ax} \right) \phi_{A} \phi_{x}
  \rangle \: .
\end{eqnarray}
This approximation already contains 
the essential ingredients to see the principles of the TH
method and the connection to the PWIA becomes clear.
Generalizations to a more general
treatment with distorted waves are obvious from the above.
The argument of the amplitude $W$ is the momentum
\begin{eqnarray} \label{QBb}
 \vec{Q}_{Bb} & = & \vec{k}_{Bb} - \frac{m_{b}}{m_{b}+m_{x}} \vec{k}_{Aa}
\end{eqnarray} 
assuming that $\vec{K}=\vec{k}_{Aa}$.
Neglecting the Fermi motion of $b$ inside the Trojan Horse,
$m_{b} \vec{k}_{Aa} / (m_{b}+m_{x})$ is the momentum of $b$ 
relative to $A$ in the initial state and $\vec{k}_{Bb}$ is the momentum of $b$
relative to $B=C+c$ in the final state. Thus $-\vec{Q}_{Bb}$
corresponds to the momentum transfer to the spectator $b$.
The amplitude $W$ describes the distribution of the transfered
momentum due to the Fermi motion.
Similarly, the momentum 
\begin{eqnarray} \label{QAa}
 \vec{Q}_{Aa} & = & \vec{k}_{Aa} - \frac{m_{A}}{m_{A}+m_{x}} \vec{k}_{Bb}
\end{eqnarray} 
in the argument of the plane wave
can be considered as the (negative) momentum transfer to nucleus $A$ 
(independent of the choice of $\vec{K}$) by the particle $x$.
In the case of a infinitely heavy nucleus $A$, equations (\ref{QBb})
and (\ref{QAa}) reduce to
\begin{equation}
 \vec{Q}_{Bb}  =  \frac{m_{b}}{m_{b}+m_{x}} \vec{k}_{a} - \vec{k}_{b} 
 = \vec{k}_{x} - \frac{m_{x}}{m_{b}+m_{x}} \vec{k}_{a}
 \: , \qquad 
  \vec{Q}_{Aa}  =  \vec{k}_{b} - \vec{k}_{a}  = - \vec{k}_{x}
\end{equation}
(cf.\ Fig.\ \ref{fig00})
and the interpretation becomes simpler.
The main task is to calculate the TH integrals (\ref{defTHI})
and to find the explicit relation of the three-body cross section
to the two-body cross section.

The plane wave approximations for $\chi^{(+)}_{Aa}$ and $\chi^{(-)}_{Bb}$
seem to be crude at first sight but the TH matrix element
(\ref{TTH})
still contains the asymptotically correct 
wave function $\Psi^{(-)}_{\rm asym}(Cc,\vec{k}_{Cc})$
for the $C+c$ system and thus the complete information on the two-body scattering
process. This is in clear contrast to the PWIA \cite{Jai70} 
where the effect of the
Coulomb barrier in $A+x$ system on the energy
dependence is not obvious.
The absorptive feature of the optical potentials is taken into account
by the surface approximation. Additionally, the distorted wave 
$\chi^{(+)}_{Aa}$ describes a scattering state
with a much higher (and constant) momentum as compared to the $A+x$ system
and only a small
part of the three-body phase space is of interest in the reaction.
The absolute cross section for the three-body process (\ref{THreac})
calculated in the plane wave approximation
might be different from the actual value but the energy dependence 
with respect to the two-body reaction is expected to be treated correctly.

\section{Trojan-Horse Integrals}
\label{THI}

The main difference between the usual expressions 
for the scattering amplitude 
and the corresponding T-matrix element (\ref{TTHgen}) in the TH method is
the appearence of the functions $U_{l}^{(\pm)}$.
They decribe the off-shell behaviour of
the two-body scattering amplitude and lead to the 
effective removal of the Coulomb barrier in the quasi-free
scattering process. This effect can only be understood if the
energy dependence of these functions is known.
In the plane-wave approximation the discussion becomes
very transparent. The functions
$U_{l}^{(\pm)}$ factorize into a momentum amplitude, Legendre polynomials
and the TH integrals
(\ref{defTHI}) which are not difficult to study.

\subsection{General form of Trojan-Horse integrals}

The basic Trojan-Horse integral in the plane wave approximation is given by
\begin{equation} \label{THIH}
 J_{l}^{(H)}(R,\eta,k,Q) = k
  \int\limits_{R}^{\infty} dr \:  H_{l}(\eta;kr)
 \:   z_{l}(Q r) 
\end{equation}
where $H_{l}$ is a Coulomb wave function $F_{l}$ or $G_{l}$ and
$z_{l}(x)=xj_{l}(x)$ is a Riccati-Bessel function.
Then the integrals $J_{l}^{(\pm)}$ are obtained from
equation (\ref{THId}). The TH integrals do not converge in the usual sense,
since $H_{l}$ and $z_{l}$ oscillate with constant
amplitude if $r$ goes to infinity. Convergence is achieved only in the
distributional sense after an integration over
$Q$ with a suitable test function. The problem is caused by the fact
that the matrix element in eq.\ (\ref{TWME}) 
contains the overlap of continuum wave functions.
A similar problem occurs in the case of Bremsstrahlung matrix elements
that have to be regularized.

However, the integral (\ref{THIH}) can be transformed into a form that allows
a numerical calculation. With the differential equation
for the Coulomb wave functions
\begin{equation}
  0  =  \frac{1}{k^{2}} \frac{d^{2}}{dr^{2}} H_{l}(kr)
 + \left[ 1 - \frac{2\eta}{kr} - \frac{l(l+1)}{k^{2}r^{2}} \right]
 H_{l}(kr) 
\end{equation}
the TH integral (\ref{THIH}) becomes
\begin{eqnarray} 
 J_{l}^{(H)}(R,\eta,k,Q) & = & 
 k  \int\limits_{R}^{\infty} dr \:  
 \left[   \frac{2\eta}{kr} + \frac{l(l+1)}{k^{2}r^{2}}
 - \frac{1}{k^{2}} \frac{d^{2}}{dr^{2}}
 \right] H_{l}(\eta;kr) \:   z_{l}(Q r) 
\end{eqnarray}
After partial integration and with help of the differential equation
for $z_{l}$ 
\begin{equation}
  0  = \frac{1}{Q^{2}} \frac{d^{2}}{dr^{2}} z_{l}(Qr)
 + \left[ 1 - \frac{l(l+1)}{Q^{2}r^{2}} \right] z_{l}(Qr) 
\end{equation}
the expression
\begin{eqnarray} \label{JH}
 \lefteqn{J_{l}^{(H)}(R,\eta,k,Q)= \frac{k}{k^{2}-Q^{2}}}
  \\ \nonumber  &  & 
   \left[ 2 \eta k  I_{l}^{(H)}(R,\eta,k,Q)
  + k  H_{l}^{\prime}(\eta;kR)   z_{l}(QR) 
  - Q  H_{l}(\eta;kR)  z_{l}^{\prime} (QR)  \right]
\end{eqnarray}
with the converging integral 
\begin{equation} \label{Iint}
 I_{l}^{(H)}(R,\eta,k,Q) = 
 \int\limits_{R}^{\infty} dr \:  H_{l}(\eta;kr) \: r^{-1} \: z_{l}(Qr) 
\end{equation} 
is obtained. The prime denotes the
differentiation with respect to the argument. 
We disregard the contribution from the upper
bound (infinity). This corresponds to the regularization procedure
mentioned above. 
Relation (\ref{JH}) cannot be used if $k= Q$ but this case
will never occur since $k<Q$ from kinematical considerations
(see Section \ref{Skincon}).
The integral (\ref{Iint}) 
can be written as
\begin{eqnarray} \label{Iint2}
 I_{l}^{(H)}(R,\eta,k,Q) & = & 
 k Q  M_{ll}^{-1}(H,\eta,k,Q) -
 \int\limits_{0}^{R} dr \:  H_{l}(\eta;kr) \: r^{-1} \: z_{l}(Qr) 
\end{eqnarray}
with the integral
\begin{equation} \label{Mdef}
  M_{l_{1}l_{2}}^{-\lambda}(H,\eta,k,Q) =  \frac{1}{kQ} 
 \int\limits_{0}^{\infty} dr \:
 H_{l_{1}}(\eta;kr) r^{-\lambda} z_{l_{2}}(Qr) \: .
\end{equation}
because $I_{l}^{(H)}$ is still finite in the limit $R \to 0$. 
The functions $M_{l_{1}l_{2}}^{-\lambda}$ are
special cases of the radial matrix elements in the
quantum mechanical theory of Coulomb excitation \cite{Ald56}.
A more general form also appears in the general theory
of transfer reaction to the continuum \cite{Bau74a}.
In the THM only 
monopole matrix elements ($\lambda =1)$ are needed. They can be
calculated explicitly for both the regular and irregular
Coulomb wave functions (see Appendix \ref{AppA} for details).
The TH integrals with the regular Coulomb functions are
given by
\begin{eqnarray} \label{MF00}
 M_{00}^{-1}(F,\eta,k,Q) & = &
   \frac{C_{0}(\eta)}{2\eta kQ} 
  \sin \zeta \: ,
 \\ 
 M_{11}^{-1}(F,\eta,k,Q) & = &
 \frac{3C_{1}(\eta)}{4\eta kQ(1+\eta^{2})} 
  \left[  \left(\frac{Q}{k}+\frac{k}{Q}\right) 
 \sin \zeta - 2 \eta  \cos \zeta  \right]
\end{eqnarray}
for $l=0,1$. The argument of the trigonometric functions is
$ \zeta = \eta \ln z_{1}$ 
with $z_{1}=(Q+k)/(Q-k)$.
The constants $C_{l}(\eta)$ are recursively defined by
\cite{Abr70}
\begin{equation} \label{Cconst}
 C_{0}(\eta) = \left( \frac{2\pi \eta}{\exp(2\pi\eta)-1}\right)^{\frac{1}{2}} \: ,
 \qquad
 C_{l}(\eta) = \frac{\left(l^{2}+\eta^{2}\right)^{\frac{1}{2}}}{l(2l+1)} 
C_{l-1}(\eta) \: .
\end{equation}
The case with the irregular Coulomb wave functions is 
slightly more intricate and leads to
\begin{eqnarray}
  M_{00}^{-1}(G,\eta,k,Q) & = & 
 \frac{C_{0}(\eta)}{2 \eta kQ} 
 \left( 1- \cos \zeta \right)
 + \frac{\pi}{2kQ C_{0}(\eta)} \left[ 1 -  S_{0} \right] \: ,
 \\ 
  M_{11}^{-1}(G,\eta,k,Q) & = & 
 \frac{ 3C_{1}(\eta)}{4\eta kQ(1+\eta^{2})}
  \left[ \left(\frac{Q}{k}+\frac{k}{Q}\right) 
 \left( 1- \cos \zeta  \right)
 -2 \eta \sin \zeta \right]
 \\ \nonumber & & 
 +  \frac{\pi}{12kQ C_{1}(\eta)} 
\left[ \left(\frac{Q}{k}+\frac{k}{Q}\right)  
\left( 1 - S_{0} \right) + 2 \eta S_{1} \right]
\end{eqnarray}
with the series
\begin{eqnarray}
 S_{0}  =  \frac{2 \eta}{\pi} \sum_{n=1}^{\infty} 
 \frac{1-z_{1}^{-n}}{n^{2}+\eta^{2}} \: ,
 \qquad
 S_{1}  =  \frac{2}{\pi} \sum_{n=1}^{\infty} \frac{nz_{1}^{-n}}{n^{2}+\eta^{2}} \: .
\end{eqnarray}
Explicit expressions for larger $l$ are obtained from the
recursion relation ($l>0$)
\begin{equation} \label{recur}
   \frac{2l+1}{2} \left( \frac{k}{Q} 
 + \frac{Q}{k} \right) M_{ll}^{-1}
 =  |l+i\eta|  M_{l-1l-1}^{-1}
 +  |l+1+i\eta| M_{l+1 l+1}^{-1}
\end{equation}
which is valid for $H_{l}=F_{l}$ and $H_{l}=G_{l}$.
It can be derived in a similar manner as in Coulomb excitation theory 
\cite{Ald56}.
However, for a numerical calculation the question of stability
of (\ref{recur}) arises.
In the case of $H_{l}=F_{l}$ only the backward recurrence is stable.
Starting with, e.g., $M_{LL}^{-1}(H,\eta,k,Q)=0$ and
$M_{L-1L-1}^{-1}(H,\eta,k,Q)=1$ for large $L \gg l$ a backward recursion
gives integrals for $l=0,1,2,\dots$ after normalizing to the known value
(\ref{MF00}).
In the case of $H_{l}=G_{l}$ the upward recurrence can be used
with the known starting values $M_{00}^{-1}(G,\eta,k,Q)$ and 
$M_{11}^{-1}(G,\eta,k,Q)$.
The remaining integral from $0$ to $R$ in equation (\ref{Iint2})
is easily performed numerically. 

When $\eta$ becomes large (and $k$ small)
it is not favorable to calculate the integral 
$ I_{l}^{(G)}(R,\eta,k,Q)$ from equation (\ref{Iint2}).
The irregular Coulomb wave functions becomes very large
for small radii $r$ and a difference of large numbers has to be
evaluated with loss of accuracy. In this case it is more convenient
to use equation (\ref{Iint}) directly because $G_{l}$ becomes small
with increasing $r>R$ very rapidly. 

\subsection{Limiting cases}

In the case of $R \to 0$
the surface contributions in eq.\ (\ref{JH})
vanish only if $H_{l}=F_{l}$. This can be seen from the
approximations \cite{Abr70}
\begin{equation} \label{Fappr}
 F_{l}(\eta;x) \to \frac{(2l+1)!C_{l}(\eta)}{(2\eta)^{l+1}}
 \sqrt{2\eta x} \: I_{2l+1}(2\sqrt{2\eta x})
 \approx  C_{l}(\eta) x^{l+1}
\end{equation}
and $z_{l}(x) \to x^{l+1}/(2l+1)!!$
for $x=kR \to 0$.
In this case
the expression for TH integral with the regular Coulomb wave function  reduces
to the simple form
\begin{equation} 
 J_{l}^{(F)}(0,\eta,k,Q)  =  \frac{2\eta k^{3} Q}{k^{2}-Q^{2}}
 M_{ll}^{-1}(F_{l},\eta,k,Q) \: .
\end{equation}
Using the approximation \cite{Abr70}
\begin{equation} \label{Gappr}
 G_{l}(\eta;x) \to \frac{2(2\eta)^{l}}{(2l+1)!C_{l}(\eta)}
 \sqrt{2\eta x} \: K_{2l+1}(2\sqrt{2\eta x})
 \approx  \frac{x^{-l}}{(2l+1)C_{l}(\eta)}
\end{equation}
for small $x$
one obtains for $H_{l}=G_{l}$ the result
\begin{equation}
  J_{l}^{(G)}(0,\eta,k,Q)  =  \frac{kQ}{k^{2}-Q^{2}}
   \left[ 2 \eta k^{2}   M_{ll}^{-1}(G_{l},\eta,k,Q)
 - \frac{(Q/k)^{l}}{(2l+1)!! C_{l}(\eta)} \right]
\end{equation}
with a remaining contribution from the surface terms.

In the special case where the transfered particle $x$ is a neutron,
the Coulomb functions $u_{l}^{(\pm)}(x)$ reduce to Hankel
functions $x[-y_{l}(x) \pm i j_{l}(x)]=\pm i x h_{l}^{(\pm)}(x)$
with the spherical Neumann and Bessel functions $y_{l}$ and $j_{l}$,
respectively \cite{Abr70}. There is no contribution from the term with
the integral $I_{l}^{(H)}$ and the TH integral simplifies to
\begin{eqnarray}
 J_{l}^{(\pm)}(R,0,k,Q) =
 \frac{\pm ik^{2}QR^{2}}{k^{2}-Q^{2}}
   \left[  k  h_{l-1}^{(\pm)}(kR)   j_{l}(QR) 
  - Q  h_{l}^{(\pm)}(kR)  j_{l-1} (QR) 
 \right] 
\end{eqnarray}
so that $J_{l}^{(+)}(R,0,k,Q)$
is $ik^{2}Q$ times the $I_{l}$ integral in the 
stripping enhancement factors $F_{l}$
of Ref. \cite{Bau74b}.
For the neutral particle, there is, for $l \neq 0$, still a barrier,
the angular momentum barrier with its $l(l+1)/r^{2}$ radial dependence. 
The enhancement is due to the Hankel functions 
$h_{l}^{(\pm)}(x)$ with a $x^{-(l+1)}$ behaviour for $x \to 0$.
The mathematical treatment of the
Coulomb functions is much more involved.

\subsection{Energy dependence of Trojan-Horse integrals}
\label{EdTHI}

By a simple change of variables it is seen that the TH integrals
(\ref{THIH}) depend only on three independent parameters. With
\begin{equation}
 \xi = \frac{k}{Q} \qquad \mbox{and} \qquad x_{0} = QR
\end{equation}
we can write
\begin{equation} 
 J_{l}^{(H)}(R,\eta,k,Q)  = J_{l, {\rm red}}^{(H)}(x_{0},\eta,\xi)
 = \xi \int\limits_{x_{0}}^{\infty} dx \:  H_{l}(\eta;\xi x)
 \:   z_{l}(x) \: ,
\end{equation} 
with reduced TH integrals that
are  functions of $x_{0}$, $\eta$, and $\xi$.
In order to investigate the behaviour of $J_{l}^{(H)}$ for $k \to 0$
or $\xi \to 0$ it is useful to study the case $x_{0}=0$, i.e. $R=0$, first.
From the explicit expressions for the integrals $M_{ll}^{-1}(F_{l})$ the
approximations
\begin{eqnarray} \label{JF0app}
 J_{0}^{(F)} (0,\eta,k,Q) = J_{0, {\rm red}}^{(F)}(0,\eta,\xi) & \to &
 - C_{0}(\eta) \xi^{2} \sin ( 2 \eta \xi ) \: ,
 \\ \label{JF1app}
 J_{1}^{(F)} (0,\eta,k,Q)  = J_{1, {\rm red}}^{(F)}(0,\eta,\xi) & \to &
 -  C_{0}(\eta) \xi^{2}
   \left[ \frac{\sin ( 2 \eta \xi)}{2 \eta \xi}
 - \cos ( 2 \eta \xi) \right] 
\end{eqnarray}
for the integrals with the regular Coulomb functions $F_{0}$ and $F_{1}$ are found.
Because the recursion relation (\ref{recur}) for $M_{ll}^{-1}$
reduces to the recursion relation of the Riccati-Bessel functions
we have the approximation
\begin{equation}
 J_{l}^{(F)} (0,\eta,k,Q)  = J_{l, {\rm red}}^{(F)}(0,\eta,\xi) \to 
 - C_{0}(\eta) \xi^{2} z_{l}(2 \eta \xi)
\end{equation}
for all $l$ in the limit $k\to 0$. The argument  $2\eta \xi = 2\eta k/Q$ 
is independent of $k$ for constant $Q$ and the entire $k$ dependence is
determined by $C_{0}(\eta)\xi^{2}$ independent of $l$. 

\begin{figure}
\includegraphics[width=16cm]{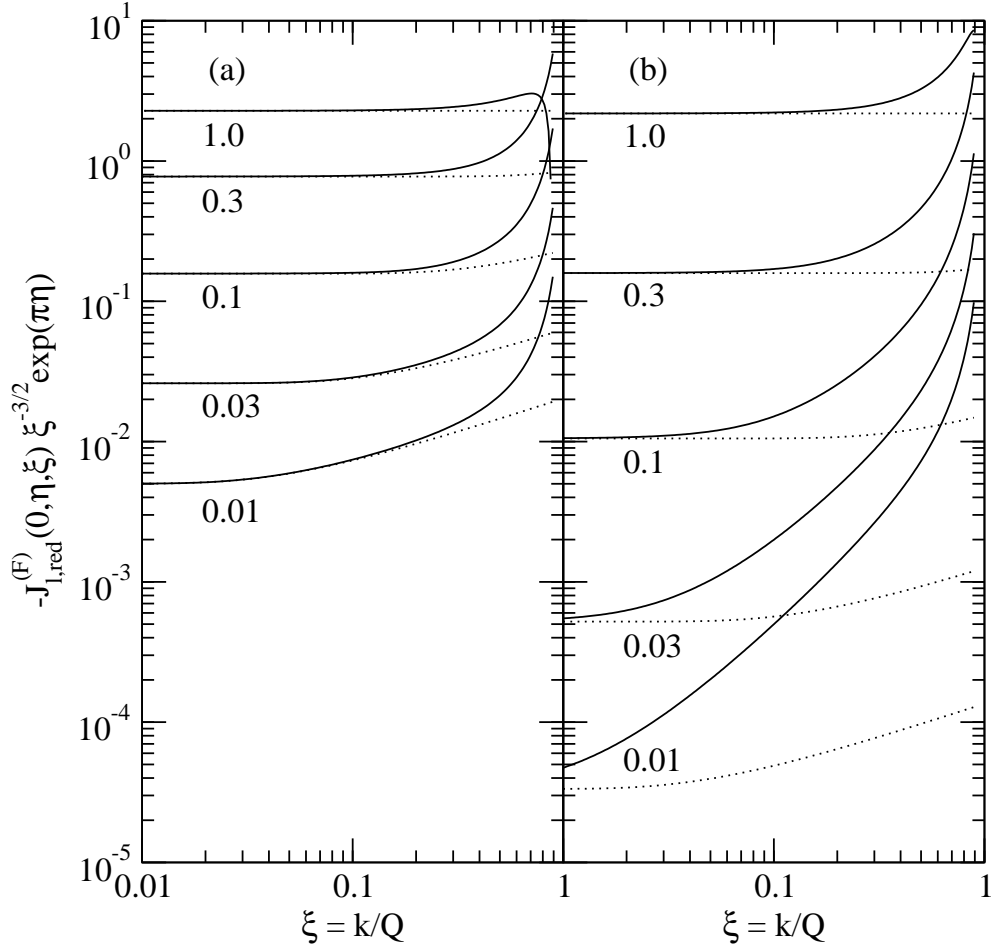}
\caption{\label{fig02} Scaled reduced Trojan-Horse integrals 
$-J_{l, {\rm red}}^{(F)}(0,\eta,\xi)\: \xi^{-\frac{3}{2}} \exp (\pi \eta) $
for parameters $\eta \xi= 0.01, \dots, 1.0$ as a function of $\xi = k/Q$
for angular momenta $l=0$ (a) and $l=1$ (b). The solid lines are the 
exact results and the dotted lines are the approximations of
Sec.\ \ref{EdTHI}.}
\end{figure}

Figure \ref{fig02} shows the dependence of the scaled reduced TH integrals
$J_{l, {\rm red}}^{(F)}(0,\eta,\xi)\: \xi^{-\frac{3}{2}}\exp(\eta\xi)$ 
with $l=0,1$ on $\xi=k/Q$ for various values of 
$\eta \xi= Z_{A}Z_{x}e^{2}\mu_{Ax}/(\hbar^{2}Q)$ 
which is a constant for given $Q$. 
The scaled TH integrals become constant for small $\xi$ as expected.
The rapid decrease of the unscaled TH integrals with decreasing $\xi$
for constant $\eta \xi$ is obvious. 
The angular momentum barrier leads to a smaller TH integral
with larger $l$ for the same parameters $\xi$ and $\eta \xi$.
This effect is more pronounced for small $\eta\xi$ than for
large $\eta\xi$ where the Coulomb barrier
dominates the $\xi$ dependence.
The approximations (\ref{JF0app}) and (\ref{JF1app}) agree very well 
with the exact results for the TH integrals with $l=0,1$ 
for large $\eta \xi$ and not too large $\xi$. The agreement
is less satisfactory for small $\eta \xi$ and larger $l$
as one may expect, since then the angular momentum
barrier is getting more important as compared to the Coulomb
barrier.

The TH integrals with the irregular
Coulomb wave functions reduce in the limit $k \to 0$ to the form
\begin{eqnarray} \label{JG0app}
 J_{0}^{(G)} (0,\eta,k,Q)  = J_{0, {\rm red}}^{(G)}(0,\eta,\xi) & \to &
  \frac{\xi}{C_{0}(\eta)}
 \left[ 1 - 2 \eta \xi \: f(2 \eta \xi) \right] \: ,
 \\ \label{JG1app}
 J_{1}^{(G)} (0,\eta,k,Q)  = J_{1, {\rm red}}^{(G)}(0,\eta,\xi) & \to &
  \frac{\xi}{C_{0}(\eta)} 
 \left[ \frac{1}{\eta\xi} 
 -  f( 2 \eta \xi) - 2 \eta \xi \: g(2 \eta \xi) \right] 
\end{eqnarray}
with functions
\begin{eqnarray}
 f(x) = \int\limits_{0}^{\infty} dt \: \frac{e^{-xt}}{t^{2}+1}
 & = & \mbox{Ci}(x) \sin x - \left[ \mbox{Si}(x)- \frac{\pi}{2} \right] \cos x
 \: ,
 \\
 g(x) = \int\limits_{0}^{\infty} dt \: \frac{t e^{-xt}}{t^{2}+1}
 & = & -\mbox{Ci}(x) \cos x - \left[ \mbox{Si}(x)- \frac{\pi}{2} \right] \sin x
\end{eqnarray}
that can be expressed in terms of the sine and cosine integrals
$\mbox{Si}(x)$ and $\mbox{Ci}(x)$ \cite{Abr70}. The expressions inside the
brackets in eqs. (\ref{JG0app}) and (\ref{JG1app}) 
are constants and the $k$-dependence of the TH integrals
with the irregular Coulomb wave functions
is determined by $\xi/C_{0}(\eta)$.

\begin{figure}
\includegraphics[width=16cm]{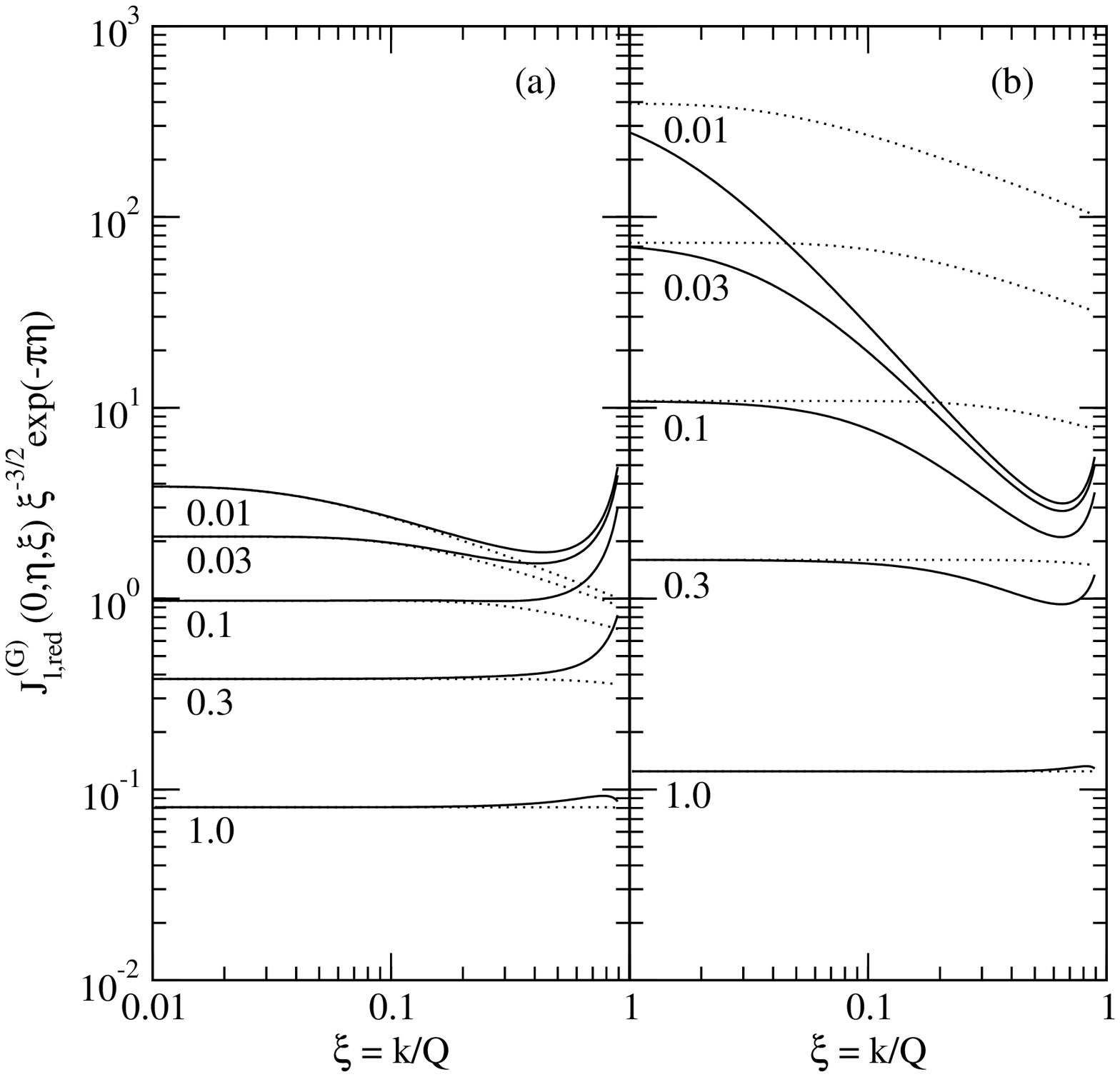}
\caption{\label{fig03} Scaled reduced Trojan-Horse integrals 
$J_{l, {\rm red}}^{(G)}(0,\eta,\xi)\: \xi^{-\frac{3}{2}} \exp(-\pi \eta)$
for parameters $\eta \xi= 0.01, \dots, 1.0$ as a function of $\xi = k/Q$
for angular momenta $l=0$ (a) and $l=1$ (b). The solid lines are the 
exact results and the dotted lines are the approximations of
Sec.\ \ref{EdTHI}.}
\end{figure}

In Figure \ref{fig03} the dependence of scaled reduced TH integrals
$J_{l, {\rm red}}^{(G)}(0,\eta,\xi)\: \xi^{-\frac{3}{2}}\exp(-\pi\eta)$ 
with $l=0,1$ on $\xi$ is shown for the same range of
the parameter $\eta \xi$ as in Fig.\ \ref{fig02}.
For small $\xi$ the scaled TH integrals again become constant.
The unscaled 
TH integrals increase dramatically for decreasing $\xi$ for constant $\eta \xi$
and for increasing $\eta \xi$ for constant $\xi$, respectively.
The effect of the angular momentum barrier is more pronounced for
small $\eta\xi$ as in the case of the TH integrals with the
regular Coulomb wave function as one may expect. The agreement between
the approximations (\ref{JG0app}) and (\ref{JG1app})
and the exact TH integrals 
shows the same trends as in the case of the  TH integrals
with the regular Coulomb wave functions.

In the case of finite $R$ the energy dependence of the TH integrals
for $\xi \to 0$
can be extracted with the help of the approximations (\ref{Fappr})
and (\ref{Gappr}) which are valid for $2 \eta \gg \xi x_{0}$.
The energy dependence of the differences
\begin{eqnarray} 
 \lefteqn{J_{l}^{(F)}(0,\eta,k,Q) - J_{l}^{(F)}(R,\eta,k,Q)
 = J_{l, {\rm red}}^{(F)}(0,\eta,\xi) 
 - J_{l, {\rm red}}^{(F)}(x_{0},\eta,\xi)}
 \\ \nonumber & &  \approx
 \frac{(2l+1)!C_{l}(\eta)\xi}{(2\eta)^{l+1}}
  \int\limits_{0}^{x_{0}} dx \:  
 \sqrt{2\eta \xi x} \: I_{2l+1}(2\sqrt{2\eta \xi x})
 \:   z_{l}(x) 
\end{eqnarray} 
and
\begin{eqnarray} 
 \lefteqn{J_{l}^{(G)}(0,\eta,k,Q) - J_{l}^{(G)}(R,\eta,k,Q)
 = J_{l, {\rm red}}^{(G)}(0,\eta,\xi) 
 - J_{l, {\rm red}}^{(G)}(x_{0},\eta,\xi)}
 \\ \nonumber & & \approx
 \frac{2(2\eta)^{l}\xi}{(2l+1)!C_{l}(\eta)}
 \int\limits_{0}^{x_{0}} dx \:  
 \sqrt{2\eta \xi x} \: K_{2l+1}(2\sqrt{2\eta \xi x})
 \:   z_{l}(x) 
\end{eqnarray} 
is entirely given by the factor in front of the integrals since
$\eta \xi = \eta k/Q$ is constant for constant $Q$. 
For large $\eta$ and small $\xi$ it is found that the
energy dependence of the finite range correction to
the TH integral with the regular and irregular
Coulomb functions is given by $\xi^{2} C_{0}(\eta)$ and
$\xi / C_{0}(\eta)$, respectively.
The finite range correction essentially leads to a change in the
absolute value of the integral but
the energy dependence of $J_{l}^{(\pm)}(R,\eta,k,Q)$
and $J_{l}^{(\pm)}(0,\eta,k,Q)$ is the same for small $k$.

When the Sommerfeld parameter $\eta$ increases,
the coefficient $C_{0}(\eta)$ becomes very small and
the TH integral with $G_{l}$ is much larger than the TH integral
with $F_{l}$. The integrals  $J_{l}^{(\pm)}$ are therefore
dominated by $J_{l}^{(G)}$ at low energies.

In the case where a neutron is transfered in reaction (\ref{THreac})
the reduced TH integrals can be written as
\begin{eqnarray}
 J_{l}^{(\pm)}(R,0,k,Q) = J_{l, {\rm red}}^{(\pm)}(x_{0},0,\xi) =
 \frac{\pm i\xi^{2}x_{0}^{2}}{\xi^{2}-1}
   \left[  \xi h_{l-1}^{(\pm)}(\xi x_{0})   j_{l}(x_{0}) 
  - h_{l}^{(\pm)}(\xi x_{0})  j_{l-1} (x_{0})  
 \right] 
\end{eqnarray}
in the variables $x_{0}=QR$ and $\xi=k/Q$ with $\eta=0$.
For $\xi \to 0$ we find the approximation
\begin{eqnarray} \label{Jlpm0app}
 J_{l, {\rm red}}^{(\pm)}(x_{0},0,\xi) \to
 \frac{(2l-1)!!}{(\xi x_{0})^{l-1}} \:  j_{l-1}(x_{0})
\end{eqnarray}
with a distinct $l$-dependence due to the centrifugal barrier.

\subsection{Simplified approximation of TH integrals}
\label{THIsap}

The dominance of the contribution with the
irregular Coulomb wave function $G_{l}$ in the
TH integrals  $J_{l}^{(\pm)}$ at low energies also motivates
a simple approximation 
in the limit $k \to 0$ that has been used in  
applications of the THM to reactions with $Ax \neq Cc$.
In this case the integral is completely determined by the
contributions at radii close to $R$ and the integrand decreases very fast
with increasing $r$.
If we assume that a general function $f(r)$ decreases
exponentially, the integral of $f(r)$ from $R$ to infinity 
can be calculated from
\begin{equation}
 \int\limits_{R}^{\infty} dr \: f(r) 
 = f(R) \int\limits_{R}^{\infty} dr \: \exp \left(-\frac{r}{\Delta R} \right)
 = \Delta R \: f(R) 
\end{equation}
with $\Delta R= -f(R)/f^{\prime}(R)$ and $\mbox{Re}(\Delta R)>0$.
In the TH integral  $J_{l}^{(\pm)}$ the appropriate
value of $\Delta R$ is determined by
\begin{equation}
 \Delta R
 \approx - \left[ \frac{kG_{l}^{\prime}(\eta;kR)}{G_{l}(\eta;kR)}
 + \frac{Q z_{l}^{\prime}(Q r)}{z_{l}(Q r)} \right]^{-1} 
\end{equation}
neglecting the small contribution of $F_{l}$ for $k\to 0$.
From the approximation (\ref{Gappr})
of the irregular Coulomb wave function for $kR \ll 2\eta$
we find 
\begin{equation}
 \Delta R \approx - \left[ \sqrt{\frac{2\eta k}{R}}
 \frac{K_{2l+1}^{\prime}(2\sqrt{2\eta k R})}{K_{2l+1}(2\sqrt{2\eta k R})}
 + \frac{Q z_{l}^{\prime}(Q r)}{z_{l}(Q r)} \right]^{-1} 
\end{equation}
independent of $k$ because $\eta k$ is constant and $Q$ is 
practically independent of $k$ for small $k$.
Applying this result to the TH integral leads to the approximation
\begin{equation}
  J_{l}^{(\pm)}(R,\eta,k,Q) = k
  \int\limits_{R}^{\infty} dr \:  u_{l}^{(\pm)}(\eta;kr)
 \:   z_{l}(Q r) 
 \approx k \Delta R \:  u_{l}^{(\pm)}(\eta;kR)  \:   z_{l}(Q R) \: .
\end{equation}
Since we are only interested in the $k$-dependence, the TH integral
can be replaced by the value of the integrand
\begin{equation}
  J_{l}^{(\pm)}(R,\eta,k,Q) 
 \to k \: R  \: u_{l}^{(\pm)}(\eta;kR)
 \:   z_{l}(Q R) 
\end{equation}
at the cutoff radius $R$ with the factor $R$ instead of $\Delta R$.
The energy dependence for small $k$
in this simple approximation
agrees with the exact result for the TH integrals.

\section{Cross Sections for Spinless Particles}

In order to establish a closer connection of the three-body and 
two-body cross sections the Trojan-Horse T-matrix element (\ref{TTH}) 
is recast in the form
\begin{equation} 
 T_{fi}^{TH}(\vec{k}_{Cc},\vec{k}_{Bb};\vec{k}_{Aa})
 = \frac{4\pi}{k_{Ax}Q_{Aa}}
  \: W(\vec{Q}_{Bb}) \: f^{TH}(\vec{Q}_{Aa}, \vec{k}_{Cc})
\end{equation}
with the Trojan-Horse scattering amplitude
\begin{eqnarray} \label{fTH}
  f^{TH}(\vec{Q}_{Aa}, \vec{k}_{Cc}) & = & 
 \frac{k_{Ax}Q_{Aa}}{4\pi}  
 \langle   \Psi^{(-)}_{Cc,{\rm asym}}(\vec{k}_{Cc},\vec{r}_{Cc})  | 
 \exp \left( i \vec{Q}_{Aa} \cdot \vec{r}_{Ax} \right) \phi_{A} \phi_{x}
  \rangle
 \\ \nonumber & = & 
  \frac{1}{2ik_{Cc}}
 \sum_{l} (2l+1)  P_{l} (\hat{k}_{Cc}\cdot\hat{Q}_{Aa})
  \sqrt{\frac{v_{Cc}}{v_{Ax}}}
 \\ \nonumber & & 
 \left[ S^{l}_{AxCc}  J_{l}^{(+)}(R,\eta_{Ax},k_{Ax},Q_{Aa})
 - \delta_{Ax Cc}  J_{l}^{(-)}(R,\eta_{Ax},k_{Ax},Q_{Aa}) \right] 
\end{eqnarray}
which resembles an amplitude for the two-body reaction 
\begin{equation} \label{Ireac}
 C+c \to A+x \: .
\end{equation}
The essential difference is the appearence of
the Trojan-Horse integrals (\ref{defTHI}).
The argument of the Legendre polynomial $P_{l}$ in equation (\ref{fTH})
corresponds to the
usual cosine of the scattering angle $\vartheta_{cm}$. 
It is convenient to decompose the TH amplitude 
\begin{eqnarray}
  f^{TH} 
 & = & f_{N}^{TH} + f_{C}^{TH} \delta_{Ax Cc} 
\end{eqnarray}
into a nuclear part and a purely Coulomb part.
The nuclear contribution 
\begin{eqnarray}
  f_{N}^{TH}  & = & 
 \frac{1}{2ik_{Cc}} 
 \sum_{l} (2l+1)  P_{l}
 (\hat{k}_{Cc}\cdot \hat{Q}_{Aa})
  \sqrt{\frac{v_{Cc}}{v_{Ax}}} 
  T^{l}_{AxCc} J_{l}^{(+)} (R,\eta_{Ax},k_{Ax},Q_{Aa})
\end{eqnarray}
with
\begin{equation}
  T^{l}_{AxCc} =  e^{i\sigma_{l}(\eta_{Ax})+i\sigma_{l}(\eta_{Cc})} 
  \left[  S^{Nl}_{AxCc}  -  \delta_{Ax Cc}  \right]  
\end{equation} 
depends on the nuclear S-matrix element
\begin{equation}
 S^{Nl}_{AxCc} = 
  e^{-i\sigma_{l}(\eta_{Ax})}  S^{l}_{AxCc} e^{-i\sigma_{l}(\eta_{Cc})} 
\end{equation}
which is obtained from the full S-matrix element by compensating
the Coulomb phase shifts $\sigma_{l}(\eta_{Cc})$ and 
$\sigma_{l}(\eta_{Ax})$ in the initial and final states, respectively.
The Coulomb contribution
\begin{eqnarray} \label{fCTH}
  f_{C}^{TH}  & = & 
  \frac{1}{k_{Cc}} 
 \sum_{l} (2l+1)  P_{l}
 (\hat{k}_{Cc}\cdot\hat{Q}_{Aa})
  e^{i\sigma_{l}(\eta_{Cc})}  J_{l}^{(F)}(R,\eta_{Ax},k_{Ax},Q_{Aa})
\end{eqnarray}
appears only if elastic two-body reactions are studied. It depends
only in the TH integrals with the regular Coulomb wave function
and is discussed in Section \ref{SfTHC}.
 
If the Trojan-Horse integrals $J_{l}^{(\pm)}$
are replaced by one the scattering amplitudes reduce to the standard
form for the reaction (\ref{Ireac}).
At low energies there are usually only  few contributions to
the nuclear scattering amplitude with small angular momenta $l$
due to the increase of centrifugal barrier with $l$ in the two-body reaction.

The cross section (\ref{d3slab}) in the laboratory system
can now be expressed in the form
\begin{eqnarray} \label{d3slabWdsTH}
 \frac{d^{3}\sigma}{dE_{C}d\Omega_{C}d\Omega_{c}} 
 & = & K_{\rm lab} \left|  W(\vec{Q}_{Bb}) \right|^{2}
  \frac{16\pi^{2}}{k_{Ax}^{2}Q_{Aa}^{2}} 
 \frac{v_{Cc}}{v_{Ax}}
 \frac{d\sigma^{TH}}{d\Omega}
\end{eqnarray}
with the two-body Trojan-Horse cross section
\begin{equation} \label{THxs}
\frac{d\sigma^{TH}}{d\Omega} =
 \frac{v_{Ax}}{v_{Cc}} \left| f^{TH}  \right|^{2} \: .
\end{equation} 
A corresponding expression holds for the c.m.\ cross section
(\ref{d3scm}) with the appropriate kinematical factor.
This result has a similar structure as in the PWIA, i.e.\
a product of a kinematical factor, a momentum distribution and
a two-body cross section. However, the momentum distribution 
$\left|  W \right|^{2}$ is not directly the ground
state momentum distribution $\left|\Phi_{a}\right|^{2}$ of the
Trojan Horse $a$ and the  Trojan-Horse cross section (\ref{THxs})
contains explicitly the off-shell effects. It is a sum
\begin{equation} \label{THxs2}
\frac{d\sigma^{TH}}{d\Omega} 
 = \left[ \frac{d\sigma^{TH}_{C}}{d\Omega} + \frac{d\sigma^{TH}_{I}}{d\Omega}
 \right] \delta_{Ax Cc}
 + \frac{d\sigma^{TH}_{N}}{d\Omega}
\end{equation} 
of a Coulomb contribution
\begin{equation}
\frac{d\sigma^{TH}_{C}}{d\Omega} =
 \left| f^{TH}_{C}  \right|^{2}
\end{equation} 
a Coulomb-nuclear interference contribution
\begin{eqnarray}
\frac{d\sigma^{TH}_{I}}{d\Omega} & = & 
  \sum_{l} \frac{2l+1}{k_{Cc}}  P_{l} (\hat{k}_{Cc}\cdot \hat{Q}_{Aa})
  \: \mbox{Re} \left[ \left(if^{TH}_{C} \right)^{\ast}
  T^{l}_{AxCc} J_{l}^{(+)}(R,\eta_{Ax},k_{Ax},Q_{Aa}) 
  \right]
\end{eqnarray} 
and a nuclear contribution
\begin{eqnarray} \label{xsTHN}
\frac{d\sigma^{TH}_{N}}{d\Omega} & = & 
 \frac{1}{4k_{Cc}^{2}} \sum_{\lambda} \sum_{l l^{\prime}} 
 (2l+1) (2l^{\prime}+1) 
 ( l \: 0 \: l^{\prime} \: 0 | \lambda \: 0 )^{2}
  P_{\lambda}(\hat{k}_{Cc}\cdot \hat{Q}_{Aa})
 \\ \nonumber & & 
 T^{l}_{AxCc} 
 T^{l^{\prime}\ast}_{AxCc} 
 J_{l}^{(+)}(R,\eta_{Ax},k_{Ax},Q_{Aa}) 
 J_{l^{\prime}}^{(-)}(R,\eta_{Ax},k_{Ax},Q_{Aa}) \: .
\end{eqnarray} 
Again, the expression for the TH cross section closely resembles the usual
c.m.\ cross section for the two-body reaction (\ref{Ireac}). The appearence of
the TH integrals $J_{l}^{(\pm)}$ accounts for the off-shell effects
and the scalar product $\hat{k}_{Cc}\cdot \hat{Q}_{Aa}$ appears
as the argument of the Legendre polynomial instead of the
cosine of the two-body c.m.\ scattering angle $\vartheta_{cm}$. 
The TH Coulomb scattering amplitude (\ref{fCTH})
replaces the  on-shell Coulomb scattering amplitude 
\begin{eqnarray} \label{fC}
 f_{C} & = &
 \frac{1}{2ik_{Cc}} 
 \sum_{l} (2l+1)  P_{l}
 (\cos\vartheta_{cm}) \left( \exp[2i\sigma_{l}(\eta_{Cc})] -1 \right)
 \\ \nonumber & = & 
- \frac{\eta_{Cc}}{2k_{Cc}\sin^{2}\frac{\vartheta_{cm}}{2}}
 \exp \left[ 2i\sigma_{0}(\eta_{Cc})
 - 2i \eta_{Cc} \ln \sin \frac{\vartheta_{cm}}{2} \right]
\end{eqnarray}
in the usual elastic two-body scattering amplitude.

The expression for the TH cross section simplifies considerably
in special cases.
If $Ax\neq Cc$ and only one partial wave $l_{Ax}$ contributes to the cross 
section the Trojan-Horse cross section is given by
\begin{equation}
 \frac{d\sigma^{TH}}{d\Omega} = 
 \frac{d\sigma_{l}}{d\Omega} (Cc \to Ax)
 \:  P_{l}(R,\eta_{Ax},k_{Ax},Q_{Aa})
\end{equation}
with the usual partial on-shell cross section
\begin{eqnarray}
\frac{d\sigma_{l}}{d\Omega} (Cc \to Ax) & = & 
 \frac{(2l+1)^{2}}{4k_{Cc}^{2}} \left| T^{l}_{AxCc} \right|^{2}
 \sum_{\lambda}  
 ( l \: 0 \: l \: 0 | \lambda \: 0 )^{2}
  P_{\lambda}(\hat{k}_{Cc}\cdot \hat{Q}_{Aa})
\end{eqnarray} 
for the two-body reaction (\ref{Ireac})
and the penetrability factor
\begin{equation} \label{pene}
 P_{l}(R,\eta_{Ax},k_{Ax},Q_{Aa})
 = \left| J_{l}^{(+)}(R,\eta_{Ax},k_{Ax},Q_{Aa}) \right|^{2} \: .
\end{equation}
In the simple approximation of Sec.\ \ref{THIsap} it is given by
\begin{equation} 
 P_{l}(R,\eta,k,Q)
 \to   k^{2} R^{2} \left[F_{l}^{2}(\eta;kR) + G_{l}^{2}(\eta;kR) \right]
 \:   z_{l}^{2}(Q R) \: .
\end{equation}
From the $k$-dependence of the TH integrals or the Coulomb wave functions
a $k^{3} \exp(2\pi\eta)$
behaviour of the penetrability factor is found for $k\to 0$.
For the transfer of a neutron the dependence of $P_{l}$ on small $k$ 
is given by $k^{2-2l}$, see eq.\ (\ref{Jlpm0app}), with an $l$-dependence 
determined by the centrifugal barrier.

With the theorem of detailed balance one obtains
the direct relation
\begin{eqnarray} \label{xslrel}
 \frac{d^{3}\sigma}{dE_{C}d\Omega_{C}d\Omega_{c}} 
 & = & K_{\rm lab} \left|  W(\vec{Q}_{Bb}) \right|^{2}
  \frac{16\pi^{2}}{k_{Cc}^{2}Q_{Aa}^{2}} 
 \frac{v_{Cc}}{v_{Ax}}
 \frac{d\sigma_{l}}{d\Omega} (Ax \to Cc)
 \:  P_{l}(R,\eta_{Ax},k_{Ax},Q_{Aa}) 
\end{eqnarray}
of the three-body cross section to the two-body cross section for partial wave $l$.
This equation shows clearly the ``parallelism'' of the $A+x$ reaction
and the Trojan-Horse reaction: The cross section (\ref{xslrel})
is proportional to the  cross section for
the $A+x$ reaction, modulated 
by the penetrability factor $P_{l}$. 
The factor $P_{l}$ is directly related to the TH
integrals (\ref{pene}).
It leads in general to an enhancement
of the higher partial waves and it contains the $\exp(2\pi \eta)$ 
factor.

A most convincing
beautiful example of this parallelism is the comparison of
neutron elastic scattering and the $(d,p)$ reaction on ${}^{15}$N 
in the same energy range of the continuum in ${}^{16}$N. There are the same
peaks in both spectra, changed in magnitude according to the factor $P_{l}$
($\eta =0$ in the neutron transfer case). Also the s-wave resonance which
appears as a destructive interference with the $l=0$ continuum shows up
nicely in the $(d,p)$ spectrum \cite{Fuc71}.

\section{Cross Sections for Particles with Spin}

For the application of the THM to a particular reaction one
has to consider that
the nuclei participating in the reactions (\ref{APreac})
and (\ref{THreac})
usually carry a spin. In this case the definition
of the relevant quantities becomes more intricate
but the general procedure remains the same as in the spinless case.
Spins of an individual particle $i$ will be denoted
in the following by $s_{i}$ with projection $\nu_{i}$.

Here, the channel spin basis in the derivation of 
the expressions is used.
In the two-particle reaction the spins $s_{A}$ and $s_{x}$
in the initial state are coupled to the channel spin $s_{Ax}$.
Similarly, in the final state $s_{C}$ and $s_{c}$ are coupled
to the channel spin $s_{Cc}$. Coupling $s_{Cc}$ with the spin
of the spectator $s_{b}$ gives the angular momentum $s_{f}$ 
of the three-body final state. The initial state of reaction
(\ref{THreac}) is characterized by the angular momentum $s_{i}$
which is obtained from coupling the spins $s_{A}$ and $s_{a}$.

The cross section of reaction (\ref{THreac}) in the laboratory system
\begin{eqnarray} 
 \frac{d^{3}\sigma}{dE_{C}d\Omega_{C}d\Omega_{c}}  & = &
 \frac{K_{\rm lab}}{(2s_{A}+1)(2s_{a}+1)} \sum_{s_{i} s_{f}}
 \sum_{\nu_{i} \nu_{f}} \sum_{s_{Cc}}
 \left| T_{fi} (\vec{k}_{Cc},\vec{k}_{Bb}, s_{Cc} s_{f} \nu_{f};
 \vec{k}_{Aa}, s_{i} \nu_{i})
 \right|^{2} 
\end{eqnarray}
is obtained by summing over  final spin states and 
averaging over initial spin states
with the three-body T-matrix element $T_{fi}$ 
that carries now both momentum and
spin indices. In the plane wave approximation it is given by
\begin{eqnarray}
 \lefteqn{T_{fi}^{TH}(\vec{k}_{Cc},\vec{q}_{Bb},s_{Cc} s_{f} \nu_{f} 
 ;\vec{q}_{Aa},s_{i} \nu_{i})}
 \\ \nonumber  & = & 
 \langle \exp \left( i \vec{q}_{Bb} \cdot \vec{r}_{Bb} \right)
  \Psi^{(-)}_{\rm asym}(Cc-b,\vec{k}_{Cc},s_{f} \nu_{f})   | V_{xb} |
 \exp \left( i \vec{q}_{Aa} \cdot \vec{r}_{Aa} \right)
 \phi_{Aa}(s_{i} \nu_{i}) \rangle
\end{eqnarray}
with the asymptotic wave function
\begin{eqnarray}
 \lefteqn{\Psi^{(-)}_{\rm asym}(Cc-b,\vec{k}_{Cc},s_{Cc},s_{f} \nu_{f})}
 \\ \nonumber & = & \sum_{\nu_{Cc} \nu_{b}}
   ( s_{Cc} \: \nu_{Cc} \: s_{b} \: \nu_{b} | s_{f} \: \nu_{f} )
 \Psi^{(-)}_{\rm asym}
 (Cc,\vec{k}_{Cc},s_{Cc} \nu_{Cc}) \phi_{b}(s_{b} \nu_{b})
\end{eqnarray}
where
\begin{eqnarray}
 \lefteqn{\Psi^{(\pm)}_{\rm asym} 
 (Cc, \vec{k}_{Cc}, s_{Cc}, \nu_{Cc} )} \\ \nonumber
 & = & \frac{4\pi}{k_{Cc}} \sum_{\alpha s_{\alpha}} \sum_{JM}
 \sum_{l_{\alpha} l_{Cc}} \frac{1}{r_{\alpha}}
 \xi^{J(\pm)}_{l_{\alpha} s_{\alpha}l_{Cc}s_{Cc}} 
 (\alpha ; k_{\alpha}  r_{\alpha})
 {\cal Y}_{JM}^{l_{\alpha}s_{\alpha}} (\alpha, \hat{r}_{\alpha})
 {\cal Z}_{JM}^{l_{Cc}s_{Cc}\nu_{Cc}\ast}(\hat{k}_{Cc})
\end{eqnarray}
for $r_{\alpha} \geq R$ and $\Psi^{(\pm)}_{\rm asym}=0$ for $r_{\alpha} < R$.
The vector spherical harmonics  
\begin{equation}
{\cal Y}_{JM}^{ls} (\alpha, \hat{r})
 = \sum_{m\nu} (l \: m \: s \: \nu | J \: M) 
 i^{l} Y_{l m}(\hat{r}) \phi_{\alpha}(s \nu)
\end{equation}
and the angular distribution functions
\begin{equation}
{\cal Z}_{JM}^{ls\nu}(\hat{k})
 = \sum_{m} (l \: m \: s \: \nu | J \: M) 
 Y_{l m}(\hat{k}) \: .
\end{equation}
are obtained by coupling 
the channel spins $s$ with the corresponding
orbital angular momenta $l$ to the total angular momentum $J$.
The radial wave functions
\begin{eqnarray}
 \lefteqn{\xi^{J(+)}_{l_{\alpha} s_{\alpha}l_{Cc}s_{Cc}} 
 (\alpha ; k_{\alpha}  r_{\alpha}) = \frac{1}{2i} 
\sqrt{\frac{v_{Cc}}{v_{\alpha}}}}
\\ \nonumber  & & 
  \left[ S^{J \alpha Cc}_{l_{\alpha} s_{\alpha} l_{Cc} s_{Cc}} 
 u_{l_{\alpha}}^{(+)}(\eta_{\alpha}; k_{\alpha}r_{\alpha})
 - \delta_{\alpha Cc} \delta_{l_{\alpha} l_{Cc}} \delta_{s_{\alpha} s_{Cc}}
 u_{l_{\alpha}}^{(-)}(\eta_{\alpha}; k_{\alpha}r_{\alpha}) \right]
\end{eqnarray}
and
\begin{equation}
 \xi^{J(-)}_{l_{\alpha} s_{\alpha}l_{Cc}s_{Cc}} 
 (\alpha ; k_{\alpha}  r_{\alpha})
 = \xi^{J(+)\ast}_{l_{\alpha} s_{\alpha}l_{Cc}s_{Cc}} 
 (\alpha ; k_{\alpha}  r_{\alpha})
\end{equation}
contain the general S-matrix elements 
$S^{J\alpha Cc}_{l_{\alpha} s_{\alpha} l_{Cc} s_{Cc}}$ for a transition
from a channel with quantum numbers $l_{Cc}s_{CC}$ in the partition $Cc$ to the
channel $l_{\alpha}s_{\alpha}$ in partition $\alpha$.

The T-matrix in the plane wave approximation
\begin{eqnarray}
 T_{fi}^{TH}(\vec{k}_{Cc},\vec{q}_{Bb},s_{Cc} s_{f} \nu_{f} 
 ;\vec{q}_{Aa},s_{i} \nu_{i}) & = & 
 \frac{4\pi}{k_{Ax}Q_{Aa}}
  \: W(\vec{Q}_{Bb}) \: {\cal F}^{TH}
(\vec{Q}_{Aa}, \vec{k}_{Cc}, s_{Cc} s_{f} \nu_{f} s_{i} \nu_{i} ) 
\end{eqnarray}
again factorizes into a form with the momentum amplitude $W$ and
the scattering amplitude
\begin{eqnarray}
  \lefteqn{{\cal F}^{TH}
(\vec{Q}_{Aa}, \vec{k}_{Cc}, s_{Cc} s_{f} \nu_{f} s_{i} \nu_{i} )=}
 \\ \nonumber & & 
 \frac{1}{2ik_{Cc}} \sum_{J} \sum_{s_{Ax}} 
 \sum_{l_{Ax} l_{Cc}} (2J+1)  {\cal P}_{l_{Ax}s_{Ax}l_{Cc}s_{Cc}}^{J}
 (\hat{k}_{Cc},\hat{Q}_{Aa},s_{f} \nu_{f} s_{i} \nu_{i})
 \sqrt{\frac{v_{Cc}}{v_{Ax}}}
 \\ \nonumber & &
 \left[ S^{J Ax Cc}_{l_{Ax} s_{Ax} l_{Cc} s_{Cc}}  
 J_{l_{Ax}}^{(+)} (R, \eta_{Ax}, k_{Ax}, Q_{Aa}) 
 - \delta_{Ax Cc} \delta_{l_{Ax} l_{Cc}} \delta_{s_{Ax} s_{Cc}}
 J_{l_{Ax}}^{(-)} (R, \eta_{Ax}, k_{Ax}, Q_{Aa}) \right]
\end{eqnarray}
which is a matrix in spin space.
The angular distribution is determined by the function
\begin{eqnarray}
  \lefteqn{{\cal P}_{l_{Ax}s_{Ax}l_{Cc}s_{Cc}}^{J}
 (\hat{k}_{Cc},\hat{Q}_{Aa},s_{f} \nu_{f} s_{i} \nu_{i})}
 \\ \nonumber 
  & = & \frac{4 \pi}{2J+1} 
  \sum_{\nu_{x} \nu_{b}} \sum_{\nu_{A} \nu_{a}} \sum_{m_{Cc} \nu_{Cc}}
  \sum_{m_{Ax} \nu_{Ax}}  \sum_{M}   
 Y_{l_{Cc}m_{Cc}}(\hat{k}_{Cc})
 Y_{l_{Ax}m_{Ax}}^{\ast}(\hat{Q}_{Aa})
 \\ \nonumber & & 
  ( s_{x} \: \nu_{x} \: s_{b} \: \nu_{b} | s_{a} \: \nu_{a} )
  ( s_{A} \: \nu_{A} \: s_{a} \: \nu_{a} | s_{i} \: \nu_{i} )
  ( s_{Cc} \: \nu_{Cc} \: s_{b} \: \nu_{b} | s_{f} \: \nu_{f} )
 \\ \nonumber & & 
  ( l_{Ax} \: m_{Ax} \: s_{Ax} \: \nu_{Ax} | J \: M )
  ( s_{A} \: \nu_{A} \: s_{x} \: \nu_{x} | s_{Ax} \: \nu_{Ax} )
  ( l_{Cc} \: m_{Cc} \: s_{Cc} \: \nu_{Cc} | J \: M )
 \\ \nonumber 
  & = & 4 \pi \sum_{m_{Ax} m_{Cc}} \sum_{jm} 
   ( s_{i} \: \nu_{i} \: j \: m | s_{f} \: \nu_{f} )
   ( l_{Cc} \: m_{Cc} \: j \: m | l_{Ax} \: m_{Ax} )
 \\ \nonumber & & 
 Y_{l_{Cc}m_{Cc}}(\hat{k}_{Cc})
 Y_{l_{Ax}m_{Ax}}^{\ast}(\hat{Q}_{Aa})
 {\cal X}_{l_{Ax}s_{Ax}l_{Cc}s_{Cc}}^{Jjs_{i}s_{f}}(s_{x},s_{b},s_{a},s_{A})
\end{eqnarray}
with
\begin{eqnarray}
 \lefteqn{{\cal X}_{l_{Ax}s_{Ax}l_{Cc}s_{Cc}}^{Jjs_{i}s_{f}}(s_{x},s_{b},s_{a},s_{A})}
 \\ \nonumber & = & 
  \nonumber   
 (-1)^{s_{Ax}-s_{x}-s_{A}-j} 
  \frac{(-1)^{s_{Ax}+J+l_{Cc}}}{\sqrt{2s_{Ax}+1}\sqrt{2l_{Ax}+1}}
  \sqrt{2s_{a}+1} \sqrt{2s_{i}+1} 
  \\ \nonumber & &  
   (2j+1)  (2s_{Ax}+1)
   \left\{ \begin{array}{ccc}
 s_{x} & s_{A} & s_{Ax} \\ s_{i} & s_{b} & s_{a} 
  \end{array} \right\}
    \left\{ \begin{array}{ccc}
 s_{Ax} & s_{b} & s_{i} \\ s_{f} & j & s_{Cc} 
  \end{array} \right\}
    \left\{ \begin{array}{ccc}
 s_{Ax} & J & l_{Ax} \\ l_{Cc} & j & s_{Cc} 
  \end{array} \right\} \: .
\end{eqnarray}
The Trojan-Horse scattering amplitude is a sum
\begin{eqnarray}
  {\cal F}^{TH} & = & 
 {\cal F}_{N}^{TH} + {\cal F}_{C}^{TH}
 \delta_{Ax Cc} 
\end{eqnarray}
of the pure Coulomb contribution
\begin{eqnarray}
  {\cal F}_{C}^{TH}  & = & 
  \frac{1}{k_{Cc}} \sum_{J}  \sum_{s_{Ax}} \sum_{l_{Ax} l_{Cc}} 
 (2J+1)  {\cal P}_{l_{Ax}s_{Ax}l_{Cc}s_{Cc}}^{J}
 (\hat{k}_{Cc},\hat{Q}_{Aa},s_{f} \nu_{f} s_{i} \nu_{i})
 \\ \nonumber & &
  e^{i\sigma_{l_{Ax}}(\eta_{Cc})} 
 J_{l_{Ax}}^{(F)} (R, \eta_{Ax}, k_{Ax}, Q_{Aa})
 \delta_{l_{Ax} l_{Cc}} \delta_{s_{Ax} s_{Cc}}
 \\ \nonumber   & = & 
    f^{TH}_{C}  \delta_{\nu_{i} \nu_{f}} \delta_{s_{i}s_{f}}
  (-1)^{s_{x}+s_{A}+s_{b}+s_{f}}
  \sqrt{2s_{Cc}+1} \sqrt{2s_{a}+1}  
   \left\{ \begin{array}{ccc}
 s_{x} & s_{A} & s_{Cc} \\ s_{f} & s_{b} & s_{a} 
  \end{array} \right\}
\end{eqnarray}
and the nuclear contribution
\begin{eqnarray}
  {\cal F}_{N}^{TH}  & = & 
 \frac{1}{2ik_{Cc}} \sum_{J} \sum_{s_{Ax}} 
 \sum_{l_{Ax} l_{Cc}} (2J+1)  {\cal P}_{l_{Ax}s_{Ax}l_{Cc}s_{Cc}}^{J}
 (\hat{k}_{Cc},\hat{Q}_{Aa},s_{f} \nu_{f} s_{i} \nu_{i})
 \\ \nonumber & &
   \sqrt{\frac{v_{Cc}}{v_{Ax}}} 
 T^{J Ax Cc}_{l_{Ax} s_{Ax} l_{Cc} s_{Cc}} 
 J_{l_{Ax}}^{(+)} (R, \eta_{Ax}, k_{Ax}, Q_{Aa})
\end{eqnarray}
with
\begin{eqnarray}
 T^{J Ax Cc}_{l_{Ax} s_{Ax} l_{Cc} s_{Cc}}  & = & 
  e^{i\sigma_{l_{Ax}}(\eta_{Ax})+i\sigma_{l_{Cc}}(\eta_{Cc})} 
    \left[  S^{J Ax Cc N}_{l_{Ax} s_{Ax} l_{Cc} s_{Cc}} 
 -  \delta_{Ax Cc} \delta_{l_{Ax} l_{Cc}} \delta_{s_{Ax} s_{Cc}} 
 \right] 
\end{eqnarray}
depending on the nuclear S-matrix element $S^{J Ax Cc N}_{l_{Ax} s_{Ax} l_{Cc} s_{Cc}}$.

The unpolarized cross section for the three-body reaction in the laboratory system
\begin{eqnarray}
 \frac{d^{3}\sigma}{dE_{C}d\Omega_{C}d\Omega_{c}}
 & = & 
 \frac{K_{\rm lab}}{(2s_{A}+1)(2s_{a}+1)} \sum_{\nu_{i} \nu_{f}} \sum_{s_{i} s_{f}}
 \sum_{s_{Cc}}
 \left| T_{fi} (\vec{k}_{Cc},\vec{k}_{Bb},s_{Cc} s_{f} \nu_{f} 
 ;\vec{k}_{Aa},s_{i} \nu_{i}) \right|^{2}
 \\ \nonumber 
 & = & K_{\rm lab} \left|  W(\vec{Q}_{Bb}) \right|^{2}
  \frac{16\pi^{2}}{k_{Ax}^{2}Q_{Aa}^{2}} 
 \frac{v_{Cc}}{v_{Ax}}
 \frac{d\sigma^{TH}}{d\Omega}
\end{eqnarray}
is obtained by a summation over the spins in the final state and
an averaging over the spins in the initial state.
As in the case of spinless particles, the TH cross section
\begin{eqnarray}
\frac{d\sigma^{TH}}{d\Omega} & = & 
 \frac{1}{(2s_{A}+1)(2s_{a}+1)}  \sum_{s_{i} s_{f}} 
\sum_{\nu_{i} \nu_{f}} \sum_{s_{Cc}}
 \frac{v_{Ax}}{v_{Cc}} \left| {\cal F}^{TH}  \right|^{2}
 \\ \nonumber 
 & = & \left[ \frac{d\sigma^{TH}_{C}}{d\Omega}
 + \frac{d\sigma^{TH}_{I}}{d\Omega} \right] 
 \delta_{AxCc}
 + \frac{d\sigma^{TH}_{N}}{d\Omega}
\end{eqnarray} 
decomposes into a pure Coulomb contribution
\begin{eqnarray}
\frac{d\sigma^{TH}_{C}}{d\Omega} & = &
 \frac{1}{(2s_{A}+1)(2s_{a}+1)} \sum_{s_{i} s_{f}}
 \sum_{\nu_{i} \nu_{f}}  \sum_{s_{Cc}}
 \left| {\cal F}^{TH}_{C} \right|^{2}
 = \left| f^{TH}_{C}  \right|^{2}
\end{eqnarray} 
an interference contribution
\begin{eqnarray}
\frac{d\sigma^{TH}_{I}}{d\Omega} & = &
  \frac{1}{(2s_{A}+1)(2s_{a}+1)} \sum_{s_{i} s_{f}}
  \sum_{\nu_{i} \nu_{f}}  \sum_{s_{Cc}}
 2 \mbox{Re} \left[ {\cal F}^{TH\ast}_{C} {\cal F}^{TH}_{N} \right] 
 \\ \nonumber & = & 
    \frac{1}{(2s_{A}+1)(2s_{x}+1)} \sum_{J}
 \sum_{l_{Cc}} \sum_{s_{Cc}}
 \frac{2J+1}{k_{Cc}} 
 \\ \nonumber & &  
  \mbox{Re} \left[  \left(  i f^{TH}_{C}\right)^{\ast}  
    T^{J Cc Cc}_{l_{Cc} s_{Cc} l_{Cc} s_{Cc}} J_{l_{Cc}}^{(+)} 
  (R, \eta_{Ax}, k_{Ax}, Q_{Aa})\right]
   P_{l_{Cc}}(\hat{k}_{Cc}\cdot\hat{Q}_{Aa})
\end{eqnarray} 
and a nuclear contribution
\begin{eqnarray}
\frac{d\sigma^{TH}_{N}}{d\Omega} & = &
  \frac{1}{(2s_{A}+1)(2s_{a}+1)} \sum_{s_{i} s_{f}}
 \sum_{\nu_{i} \nu_{f}} \sum_{s_{Cc}}
 \frac{v_{Ax}}{v_{Cc}} \left| {\cal F}^{TH}_{N}  \right|^{2}
 \\ \nonumber & = & 
    \frac{1}{(2s_{A}+1)(2s_{x}+1)} 
  \frac{1}{4k_{Cc}^{2}} 
    \sum_{\lambda}
   \sum_{J J^{\prime}}  \sum_{s_{Ax} s_{Cc}}  \sum_{l_{Ax} l_{Cc}}
 \sum_{l_{Ax}^{\prime} l_{Cc}^{\prime}}    
  \\ \nonumber & &  
 (-1)^{s_{Ax}-s_{Cc}} 
 Z_{l_{Ax}l_{Ax}^{\prime}}^{JJ^{\prime}}(\lambda s_{Ax})
 Z_{l_{Cc}l_{Cc}^{\prime}}^{JJ^{\prime}}(\lambda s_{Cc})
     P_{\lambda}(\hat{Q}_{Aa}\cdot\hat{k}_{Cc})
 \\ \nonumber & & 
  T^{J Ax Cc}_{l_{Ax} s_{Ax} l_{Cc} s_{Cc}} 
  T^{J^{\prime}Ax Cc \ast}_{l_{Ax}^{\prime} s_{Ax} l_{Cc}^{\prime} s_{Cc}} 
  J_{l_{Ax}}^{(+)} (R, \eta_{Ax}, k_{Ax}, Q_{Aa})
  J_{l_{Ax}^{\prime}}^{(-)} (R, \eta_{Ax}, k_{Ax}, Q_{Aa})
\end{eqnarray}
with factors
\begin{eqnarray} 
 Z_{ll^{\prime}}^{JJ^{\prime}}(\lambda s) & = & 
  \sqrt{(2J+1)(2J^{\prime}+1)(2l+1)(2l^{\prime}+1)}
   \:  ( l \: 0 \: l^{\prime} \: 0 | \lambda \: 0 )
   \left\{ \begin{array}{ccc}
  l & \lambda & l^{\prime} \\ J^{\prime} & s & J
   \end{array} \right\}
\end{eqnarray}
for the angular momentum coupling. The Coulomb contribution is the
same as in the case of spinless particles. Notice the change from
$s_{a}$ to $s_{x}$ in the spin degeneracy factors.

If only one orbital angular momentum 
$l=l_{Ax} = l_{Ax}^{\prime}$ contributes in the $Ax$ partition
to the inelastic two-body reaction $A+x \to C+c$, the TH cross section 
\begin{equation} \label{dsTHspin}
 \frac{d\sigma^{TH}}{d\Omega} = 
 \frac{(2s_{C}+1)(2s_{c}+1)}{(2s_{A}+1)(2s_{x}+1)}
 \frac{d\sigma_{l}}{d\Omega} (Cc \to Ax)
 \:  P_{l}(R,\eta_{Ax},k_{Ax},Q_{Aa})
\end{equation}
is directly related to the usual on-shell
cross section
\begin{eqnarray}
\frac{d\sigma_{l}}{d\Omega} (Cc \to Ax) & = &
    \frac{1}{(2s_{C}+1)(2s_{c}+1)} 
  \frac{1}{4k_{Cc}^{2}} 
    \sum_{\lambda}
   \sum_{J J^{\prime}}  \sum_{s_{Ax} s_{Cc}}  \sum_{l_{Cc} l_{Cc}^{\prime}}    
  \\ \nonumber & &  
 (-1)^{s_{Ax}-s_{Cc}} 
 Z_{ll}^{JJ^{\prime}}(\lambda s_{Ax})
 Z_{l_{Cc}l_{Cc}^{\prime}}^{JJ^{\prime}}(\lambda s_{Cc})
     P_{\lambda}(\hat{Q}_{Aa}\cdot\hat{k}_{Cc})
 \\ \nonumber & & 
  T^{J Ax Cc}_{l s_{Ax} l_{Cc} s_{Cc}} 
  T^{J^{\prime} Ax Cc \ast}_{l s_{Ax} l_{Cc}^{\prime} s_{Cc}} 
\end{eqnarray}
with the penetrability factor (\ref{pene}). In (\ref{dsTHspin}) 
the only difference
to the case with spinless particles is the appearance of
the spin degeneracy factors. Applying the theorem of
detailed balance the simple relation
\begin{eqnarray}
 \frac{d^{3}\sigma}{dE_{C}d\Omega_{C}d\Omega_{c}}
 & = & K_{\rm lab} \left|  W(\vec{Q}_{Bb}) \right|^{2}
  \frac{16\pi^{2}}{k_{Cc}^{2}Q_{Aa}^{2}} 
 \frac{v_{Cc}}{v_{Ax}}
 \frac{d\sigma_{l}}{d\Omega} (Ax \to Cc)
 \:  P_{l}(R,\eta_{Ax},k_{Ax},Q_{Aa})
\end{eqnarray}
is found which is the same as in the spinless case. 
The increase of the factor $P_{l}$ at small $k_{Ax}$ 
compensates the decrease of the two-body cross section
$\frac{d\sigma_{l}}{d\Omega} (Ax \to Cc)$.

\section{Trojan-Horse Coulomb Scattering Amplitude}
\label{SfTHC}

In case of an inelastic two-body reaction
only the nuclear contribution to the TH scattering amplitude is
relevant. For the elastic scattering with $Ax=Cc$ also the
TH Coulomb scattering amplitude (\ref{fCTH}) contributes to the
total TH scattering amplitude. 
It depends only on the TH integrals $J_{l}^{(F)}$
with the regular Coulomb wave functions.
The summation over $l$ in eq.\ (\ref{fCTH}) poses no serious problem 
but it is convenient
to reformulate the expression as in the case of the TH integrals
in order to see the reduction of the Coulomb barrier
more clearly. For that purpose the TH Coulomb scattering amplitude
is written as a difference
\begin{equation}
 f_{C}^{TH} = f_{C0}^{TH} - f_{CR}^{TH}
\end{equation}
of a contribution (suppressing the indices of the momenta)
\begin{eqnarray}  \label{fC0TH}
  f_{C0}^{TH}  & = & 
  \frac{2\eta k^{2} Q}{k^{2}-Q^{2}}
 \sum_{l} (2l+1)  P_{l}  (\hat{k}\cdot\hat{Q})
  e^{i\sigma_{l}(\eta)}  M_{ll}^{-1}(F_{l},\eta,k,Q)
\end{eqnarray}
where the cutoff radius $R$ is set to zero
and a finite range contribution
\begin{eqnarray} \label{fCRTH}
  f_{CR}^{TH}  & = & 
    \sum_{l} (2l+1)  P_{l}
 (\hat{k}\cdot\hat{Q})  e^{i\sigma_{l}(\eta)}  L_{l}(R,\eta,k,Q)
\end{eqnarray}
with the integrals 
\begin{equation}
 L_{l}(R,\eta,k,Q) = 
  \int\limits_{0}^{R} dr \: F_{l}(\eta;kr) \: z_{l}(Q r) \: .
\end{equation} 
These are easily calculated numerically and decrease rapidly with
increasing $l$ leading to a rapid convergence of the sum (\ref{fCRTH}). 
Similar to relation (\ref{fTH}) for the full TH 
scattering amplitude, the contribution (\ref{fC0TH}) to
the TH Coulomb scattering amplitude
can be expressed as a matrix element
\begin{eqnarray}
 f_{C0}^{TH} & = & \frac{kQ}{4\pi} 
 \langle  \Psi^{(-)}_{\rm Coul}(\vec{k})  |
 \exp \left( i \vec{Q} \cdot \vec{r} \right) \rangle
\end{eqnarray}
with the pure Coulomb scattering wave function
\begin{equation} \label{cswf}
 \Psi^{(\pm)}_{\rm Coul}(\vec{k})
 = e^{-\frac{\pi}{2}\eta} \Gamma(1\pm i\eta)
 \exp (i\vec{k} \cdot \vec{r})
 {}_{1}F_{1}(\mp i\eta,1;\pm i[kr\mp \vec{k}\cdot \vec{r}])
\end{equation}
where ${}_{1}F_{1}$ denotes a confluent hypergeometric
function. This form allows an analytical calculation
with the result
\begin{eqnarray} \label{fC0THres}
 f_{C0}^{TH} & = &
 -  \frac{2\eta k^{2}Q}{Q^{2}-k^{2}}
  \frac{C_{0}(\eta)}{(\vec{Q}-\vec{k})^{2}}
 \exp \left\{ i\sigma_{0}(\eta) + i \eta \ln
  \left[\frac{Q^{2}-k^{2}}{(\vec{Q}-\vec{k})^{2}}\right] \right\}
\end{eqnarray}
as explained in Appendix \ref{AppB}. Contrary to the
ususal Coulomb scattering amplitude (\ref{fC}) there is no divergence
for $k \to 0$ since $Q$ remains finite.
In this limit, the energy dependence of the TH Coulomb scattering
amplitude is given by $ k C_{0}(\eta)$.
The apperance of the $C_{0}(\eta)$ factor in
the amplitude causes a strong reduction at small $k$.

\section{Applications of the Trojan-Horse Method}
\label{Sappl}

Several reactions have been studied with the TH method recently.
They are listed in Table~\ref{tab1} with ${}^{2}$H and 
${}^{6}$Li ($=\alpha +$d)
as typical ``Trojan Horses''.
These nuclei allow to study the transfer of
protons, neutrons, deuterons and $\alpha$-particles, which covers most
of the cases of astrophysical interest for the two-body reaction.

\begin{table}[t]
\caption{\label{tab1} Projectile energy $E_{\rm pro}$ and corresponding 
quasi-free energy $E_{Ax}^{qf}$ for pairs of two-body and Trojan-Horse 
reactions.}
\begin{tabular}{ccccc}
\hline \hline
 two-body & Trojan-Horse & $E_{\rm pro}$ & $E_{Ax}^{qf}$ & Ref. \\
 reaction & reaction &  [MeV] &  [MeV] & \\
\hline
 ${}^{2}$H(${}^{6}$Li,${}^{4}$He)${}^{4}$He &
 ${}^{6}$Li(${}^{6}$Li,${}^{4}$He${}^{4}$He)${}^{4}$He &
 6.0 & 0.029 & \cite{Mus01,Spi01,Che96} \\
 ${}^{7}$Li(p,${}^{4}$He)${}^{4}$He &
 ${}^{2}$H(${}^{7}$Li,${}^{4}$He${}^{4}$He)n &
 19.0 - 21.0 & 0.161 - 0.412 & \cite{Lat01,Ali00,Spi99,Cal97} \\
 ${}^{12}$C(${}^{4}$He,${}^{4}$He)${}^{12}$C &
 ${}^{6}$Li(${}^{12}$C,${}^{4}$He${}^{12}$C)${}^{2}$H &
 12.0 - 18.0 & 1.527 - 3.027 & \cite{Spi00,Pel00} \\
 ${}^{3}$He(${}^{2}$H,p)${}^{4}$He &
 ${}^{6}$Li(${}^{3}$He,p${}^{4}$He)${}^{4}$He &
 5.8 & 0.845 & \cite{Mus02} \\
 ${}^{6}$Li(p,${}^{3}$He)${}^{4}$He &
 ${}^{2}$H(${}^{6}$Li,${}^{3}$He${}^{4}$He)n & 
 25.0 & 1.362 & \cite{Tum02} \\
\hline \hline
\end{tabular}
\end{table}

\subsection{Kinematical Conditions}
\label{Skincon}

The Trojan Horses employed so far have a dominant s-wave configuration
in their gound state. 
Their momentum distribution $W(\vec{Q}_{Bb})$ has a maximum around
zero. Correspondingly, the equation
\begin{equation} 
 \vec{Q}_{Bb} = 0
\end{equation} 
defines the so-called quasi-free condition. In this region of the
three-body phase space
the cross section for the quasi-free reaction will reach a maximum.
From this condition the corresponding quasi-free c.m.\ energy
\begin{equation} \label{erel}
 E_{Ax}^{qf} = E_{Aa} \left( 1 - \frac{\mu_{Aa}}{\mu_{Bb}}
 \frac{\mu_{bx}^{2}}{m_{x}^{2}} \right) - \varepsilon_{a}
\end{equation}
in the initial channel of the two-body reaction  (\ref{APreac})
is derived  from energy conservation (\ref{ec3}) 
assuming the plane wave approximation.
It is obvious that even with a large c.m.\ energy $E_{Aa}$
in the entrance channel of the three-body reaction (\ref{THreac})
a small energy $E_{Ax}$ 
can be reached if a suitable Trojan Horse $a$ is chosen.
This is confirmed in Table~\ref{tab1} where the projectile energy
in the laboratory system and the quasi-free energy are shown
for several reactions.

The relation between $E_{Ax}^{qf}$ and $E_{Aa}$ is purely 
a kinematical consequence. It is not related to the Fermi motion
of particle $x$ inside the Trojan Horse $a$ which would involve
a dependence on the width of the momentum amplitude.
However, in an actual experiment a cutoff in the momentum
transfer $\vec{Q}_{Bb}$ is chosen to select the region where the
quasi-free process dominates the cross section. This procedure limits 
the range in energies $E_{Ax}$ that are within reach in the experiment
for a chosen projectile energy $E_{\rm pro}$.

In case of the quasi-free condition, all nuclei in the final state
of reaction (\ref{THreac}) are emitted in the same plane.
The momentum of the spectator $b$ is in beam direction which makes it 
difficult to detect $b$ in the experiment.
In the laboratory system the nuclei $C$ and $c$ are emitted under
angles $\vartheta_{C}$ and $\vartheta_{c}$ on opposite sides of the
beam axis. If the scattering angle $\vartheta_{cm}$ in the c.m.
system of the two-body reaction is given, the angles
$\vartheta_{C}$, $\vartheta_{c}$, the so-called
quasi-free angles, are completely specified  for
a fixed beam energy from kinematical considerations.

As a consequence, the quasi-free condition 
determines the setup of the experiment. If a particular two-body
reaction (\ref{APreac}) is to be studied close to a c.m.\ energy $E_{Ax}$
and if the Trojan Horse
$a$ is selected, then from equation (\ref{erel}) the appropriate
beam energy can be extracted. 
Since the momentum amplitude $W$ has a finite width
it is possible to study the two-body reaction in a certain energy window
around $E_{Ax}$. The c.m.\ scattering angle $\vartheta_{cm}$ determines
the arrangement of the detectors close the pair of quasi-free angles.

When $E_{Bb}$ is zero in (\ref{ec3}) 
the maximum energy 
\begin{equation}
 E_{Ax}^{\rm max} = E_{Aa} - \varepsilon_{a}
\end{equation}
in the two-body system
is reached for a fixed c.m.\ energy $E_{Aa}$ in the entrance channel
of the three-body reaction. In this case $\vec{k}_{Bb}=0$ and
$\vec{Q}_{Aa} = \vec{k}_{Aa}$. The momenta of all nuclei $C$, $c$,
and $b$ in the final state
are parallel to the beam momentum in the laboratory system.
Since $\varepsilon_{a}>0$ it follows that 
the relation
\begin{equation} \label{kQcomp}
 k_{Ax} < \sqrt{\frac{\mu_{Ax}}{\mu_{Aa}}} Q_{Aa} < Q_{Aa}
\end{equation}
holds in all kinematically allowed regions of the phase space.

\subsection{Threshold behaviour of cross sections}

The energy dependence of the two-body cross section 
\begin{eqnarray}
 \lefteqn{\frac{d\sigma}{d\Omega} (Ax \to Cc) =}
 \\ \nonumber & & 
 \frac{1}{4k_{Ax}^{2}} \sum_{\lambda} \sum_{l l^{\prime}} 
 (2l+1) (2l^{\prime}+1) 
 ( l \: 0 \: l^{\prime} \: 0 | \lambda \: 0 )^{2}
  P_{\lambda}(\cos \vartheta)
 S^{l}_{AxCc} 
 S^{l^{\prime}\ast}_{AxCc} 
\end{eqnarray} 
for the inelastic $A+x \to C+c$ reaction above the reaction threshold
is governed by the $1/k_{Ax}^{2}$ factor  and 
energy dependence 
\begin{equation}
 S^{l}_{AxCc} \propto \exp(-\pi \eta_{Ax}) 
\end{equation}
of the relevant S-matrix element.
The Coulomb barrier leads to a strong suppression of the
the cross section 
\begin{equation}
 \frac{d\sigma}{d\Omega} (Ax \to Cc) \propto 
  k_{Ax}^{-2} \exp(-2 \pi \eta_{Ax})
\end{equation} 
for $k_{Ax} \to 0$ due to the decreasing exponential factor.
This behaviour motivates the introduction of the astrophysical S-factor
(\ref{Sfac}). In the TH cross section $\frac{d\sigma^{TH}}{d\Omega}$, 
that appears in eq.\ (\ref{d3slabWdsTH}), the factor $k_{Ax}^{-2}$
is replaced with $k_{Cc}^{-2}$ and the TH integrals $J_{l}^{(+)}$ appear.
Their energy dependence for small $k_{Ax}$ is determined by
$k_{Ax}/C_{0}(\eta_{Ax})\approx k_{Ax}\exp(\pi \eta_{Ax})/ \sqrt{2\pi\eta_{Ax}}$ 
from the contribution of the irregular Coulomb wave function.
This leads to a $k_{Ax}$ dependence of the
three-body cross section (\ref{d3slabWdsTH}) according to
\begin{equation} 
 \frac{d^{3}\sigma}{dE_{C}d\Omega_{C}d\Omega_{c}} 
 \propto k_{Ax}^{-2} v_{Ax}^{-1} \exp(-2\pi \eta_{Ax})
  k_{Ax}^{2} \frac{\exp(2\pi \eta_{Ax})}{2\pi\eta_{Ax}}
 =  \left(2\pi \eta_{Ax}v_{Ax}\right)^{-1} = \mbox{const.}
\end{equation}
in the lowest order of $k_{Ax}$;
cf.\ also eq. (\ref{xslrel}) with the $k_{Ax}$ dependence of the
penetrability factor $P_{l} \propto k_{Ax}^{3} \exp(2\pi\eta_{Ax})$.
As a result the cross section does not vanish at the treshold
but takes on a finite value. Of course, the same considerations 
apply to the case when the spins of the particles are considered.
Also in the case of the transfer of neutron,
like in a (d,p) stripping reaction, it is well known
that the cross section is finite
at the threshold $E_{n}=0$ \cite{Bau84,Bau76}. 
The reason is the same as in the case of
charged particles.

\subsection{Extraction of astrophysical S-factors}

In an actual TH experiment the measured cross section of the
three-body reaction depends on the geometry 
of the setup, the chosen Trojan Horse, and the energy of the projectile.
The differential three-body cross section (\ref{d3slabWdsTH}) 
or (\ref{xslrel})
can be projected onto a simple cross section
\begin{equation} \label{xsexp}
 \frac{d\sigma}{dE} = \int dE_{C} d\Omega_{C} d\Omega_{c}
 \frac{d^{3}\sigma}{dE_{C}d\Omega_{C}d\Omega_{c}}
 \delta(E_{Ax}-E) F(E_{C},\Omega_{C}, \Omega_{c})
\end{equation}
depending on the $A+x$ c.m.\ energy $E$.
The efficiency function $F$ takes cut-offs in particle 
energies, momenta (e.g.\ $\vec{Q}_{Bb}$), the detector geometry etc.\ into account.
The experimental spectrum (\ref{xsexp}) can be compared
to a corresponding theoretical cross section from a simulation
of the experiment assuming a certain energy dependence 
of the relevant on-shell S-matrix elements of the two-body reaction,
e.g.\ from a R-matrix parametrization.
By a variation of the parameters, the best fit to the experiment
is found and the on-shell two-body cross section can
be calculated. 
If there is only a contribution of one partial wave,
the procedure becomes simpler.  
The ratio of the measured cross section (\ref{xsexp}) to the 
corresponding simulated
cross section directly gives the energy dependence of the
S-factor relative to the energy dependence assumed
in the theoretical S-factor of the two-body reaction.
Due to the uncertainties in the description of the reaction mechanism
it is expected to be difficult to extract absolute values of the cross
section. So it is better to normalize the cross section
to results from direct experiments at higher energies.
However, we expect that
the energy dependence  of the cross section (S-factor) can be
extracted much more reliably. 
For recent applications of the TH method with detailed information
on the experimental realisation we refer to 
the references \cite{Mus01,Spi01,Che96,Lat01,Ali00,Spi99,Cal97}.

\subsection{Elastic scattering with the Trojan-Horse method}

The main aim for applying the TH method is the extraction
of the energy dependence of cross sections or astrophysical S-factors
for inelastic two-body reactions. But also the indirect investigation
of elasting two-body scattering $A+x \to A+x$ can be rewarding. In the direct 
two-body scattering process the cross section is dominated by the contribution
of the Coulomb scattering amplitude (\ref{fC}) at low energies that diverges
with $k_{Ax}^{-2}$. By way of contrast, the TH Coulomb scattering amplitude
(\ref{fCTH}) vanishes for $k_{Ax} \to 0$ due to the appearance
of the TH integrals with the regular Coulomb wave function $F_{l}$.
The nuclear contribution to the TH cross section (\ref{THxs2})
becomes dominant because of the TH integrals with the
irregular Coulomb wave functions in eq.\ (\ref{xsTHN}). This allows the study of
nuclear effects in the scattering at small energies, e.g.\ the influence of
sub-threshold or low energy resonances. First attempt in this direction were made
in recent experiments \cite{Spi00,Pel00}.

\section{Summary and Outlook}

In this paper the basic theory of the Trojan-Horse method was
developed starting from a distorted wave Born approximation
of the T-matrix element. The essential surface approximation
allows to find the relation between the cross section of the
three-body reaction and the S-matrix elements of the
astrophysically relevant two-body reaction. In the 
modified plane wave approximation
the relation between the three-body and two-body cross sections
becomes very transparent. The three-body cross section is
a product of a kinematical factor, a momentum distribution and
an off-shell two-body cross section. 
Off-shell effects are expressed in terms
of so-called TH integrals that were studied in detail. Their
energy dependence leads to a finite cross section of the
three-body reaction at the threshold of the two-body reaction
without the suppression by the Coulomb barrier. This allows to
extract the energy dependence of astrophysical cross sections
from the three-body breakup reaction to very low energies
without the problems of electron screening and extremely low
cross section. A comparison of results for S-factors from direct and
indirect experiments can improve the information on the electron 
screening effect. However, dedicated Trojan-Horse experiments 
are necessary in order
to achieve a precision comparable to direct measurements.

The validity of the Trojan-Horse method can be tested by
comparing the cross sections extracted from the indirect
experiment with results from direct measurements of well studied
reactions.
In principle it is possible to assess systematic uncertainties 
of the Trojan-Horse method by studying various combinations
of projectile energies, spectators in the Trojan Horse
and scattering angles.
Furthermore, different theoretical approximations can be compared,
e.g. full DWBA calculations with and without the surface approximation
and simpler modified plane wave approximations.

One may also envisage applications of the Trojan-Horse method
to exotic nuclear beams. 
An unstable projectile hits a Trojan Horse target allowing to
study specific reactions on exotic nuclei.
A study of low-energy elastic scattering with the Trojan-Horse
method opens another application which can lead to improved
information relevant to the theoretical description of
nuclear reactions at low energies.

\acknowledgments

We gratefully acknowledge the hospitality of Claudio
Spitaleri and his group in Catania where many of the
ideas were shaped. 
This work was supported by the U.S.\ National Science
Foundation grant No.\ PHY-0070911.

\appendix

\section{Analytical Calculation of Radial Integrals}
\label{AppA}

The radial integrals 
\begin{equation} \label{MllH}
  M_{ll}^{-1}(H,\eta,k,Q) =  \frac{1}{kQ} 
 \int\limits_{0}^{\infty} dr \:
  H_{l}(\eta;kr) r^{-1} z_{l}(Qr) 
\end{equation}
appearing in equation (\ref{Iint2}) can be calculated explicitly
by using the integral representation
\begin{eqnarray}
 F_{l} (\eta;kr) + i G_{l} (\eta;kr)
 & = & \frac{i(kr)^{l+1}}{(2l+1)!C_{l}(\eta)}
 \int\limits_{i}^{i+\infty} ds \:  e^{-krs} (s-i)^{l-i\eta}
 (s+i)^{l+i\eta}
\end{eqnarray}
of the combined Coulomb function $H_{l}= F_{l}+iG_{l}$.
This form has been obtained
from the representation in \cite{Abr70} by the substitution  $t=kr(s-i)$.
The integration path in the variable $s$ is a parallel to the real axis
in the complex plane.
Since the recursion relation (\ref{recur}) connects integrals of three
successive $l$-values, only the basic integrals with $l=0$ and $l=1$
are needed for the explicit calculation of $M_{ll}^{-1}$ for all $l$.
With $z_{0}(x)= \sin x $ and $z_{1}(x) = \sin x /x - \cos x$
the integration over the radial coordinate $r$ is easily performed
and one finds
\begin{eqnarray}
 M_{00}^{-1}(H,\eta,k,Q) 
  & = &  \frac{i}{C_{0}(\eta)}  \int\limits_{i}^{i+\infty} ds \:  
 \frac{(s+i)^{i\eta}(s-i)^{-i\eta}}{k^{2}s^{2}+Q^{2}} 
\end{eqnarray}
and
\begin{eqnarray}
 M_{11}^{-1}(H,\eta,k,Q) 
 & = &  \frac{ikQ}{3C_{1}(\eta)}  \int\limits_{i}^{i+\infty} ds \:  
 \frac{(s+i)^{1+i\eta}(s-i)^{1-i\eta}}{(k^{2}s^{2}+Q^{2})^{2}} \: .
\end{eqnarray}
The substitution $s = i (t+1)/(t-1)$
leads to the expressions
\begin{eqnarray} \label{M00H0}
 M_{00}^{-1}(H,\eta,k,Q) & = & 
  \frac{-2}{C_{0}(\eta)(Q^{2}-k^{2})} I_{0}(\eta ,z) \: ,
 \\ 
 M_{11}^{-1}(H,\eta,k,Q) & = & 
  \frac{4kQ}{3C_{1}(\eta)(Q^{2}-k^{2})^{2}}  \frac{d}{dz} I_{0}(\eta, z)
 \end{eqnarray}
with the integral
\begin{equation} \label{fint}
 I_{0}(\eta ,z) = \int\limits_{C} dt \:  \frac{t^{i\eta}}{t^{2}-2tz+1} \: .
\end{equation}
The contour $C$ is the straight line from $1$ to $1+i\infty$ parallel to
the imaginary axis in the complex plane. 
The integral depends on the variable
\begin{equation}
 z = \frac{Q^{2}+k^{2}}{Q^{2}-k^{2}} 
\end{equation}
which is always larger than 1 for the conditions of the TH method.
The integrand has two poles 
at $z_{1}=z+\sqrt{z^{2}-1}>1$ and $z_{2}=z-\sqrt{z^{2}-1}<1$
on the real axis. In the next step the path of
integration is deformed to lie on the real axis and the integrand is broken
into partial fractions. This yields
\begin{eqnarray}
 I_{0}(\eta ,z) & = & \int\limits_{1}^{\infty} dt \:  \frac{t^{i\eta}}{t^{2}-2tz+1}
 - \pi i \: \mbox{Res}_{z_{1}} \frac{t^{i\eta}}{t^{2}-2tz+1}
 \\ \nonumber & = & 
 \frac{1}{z_{1}-z_{2}}
 \left[ I_{1} - I_{2} - \pi i  z_{1}^{i\eta} \right]
\end{eqnarray}
with the contribution of the residue at the pole $z_{1}$.
The remaining integrals
\begin{eqnarray}
 I_{i} & = & \int\limits_{1}^{\infty} dt \: t^{i\eta}
 \left[ \frac{1}{t-z_{i}} - \frac{1}{t} \right]
\end{eqnarray}
for $i=1,2$ are evaluated with the help of the geometric series and the
relation \cite{Abr70}
\begin{equation}
 \frac{\pi\eta}{\tanh(\pi \eta)} =  1 + 2\eta^{2}
\sum_{n=1}^{\infty} \frac{1}{n^{2}+\eta^{2}} \: .
\end{equation}
Collecting all contributions we find
\begin{eqnarray} \label{I0z}
 I_{0}(\eta ,z) & = & 
  \frac{i}{2\eta \sqrt{z^{2}-1}}
 \left[ C_{0}^{2}(\eta)   \left( z_{1}^{i\eta} -1 \right)
 -2\eta^{2} \sum_{n=1}^{\infty}   \frac{z_{1}^{-n}-1}{n^{2}+\eta^{2}}
 - \pi \eta \right]
\end{eqnarray}
for the fundamental integral (\ref{fint}).
Separating real and imaginary parts, the
explicit expressions of the integrals  $M_{00}^{-1}(H_{0},\eta,k,Q)$
and $M_{00}^{-1}(G_{0},\eta,k,Q)$ are found with eq.\ (\ref{M00H0}).
In the case $l=1$ the derivative of $I_{0}(z)$ with respect to $z$ has to be 
calculated first.

\section{Trojan-Horse Coulomb Scattering Amplitude}
\label{AppB}

Recalling the procedure to derive the radial integrals
(\ref{MllH}), the contribution $f_{C0}^{TH}$ to the Coulomb scattering amplitude
can be written as
\begin{eqnarray}
 f_{C0}^{TH} & = & \frac{\eta k^{2}Q}{2\pi(k^{2}-Q^{2})}
 \langle  \Psi^{(-)}_{\rm Coul}(\vec{k})  |  r^{-1} |
 \exp \left( i \vec{Q} \cdot \vec{r} \right) \rangle \: .
\end{eqnarray}
by considering the Schr\"{o}dinger equations for the 
Coulomb scattering wave $\Psi^{(\pm)}_{\rm Coul}$ and the plane wave.
The matrix element
can be evaluated by employing the integral representation
of the confluent hypergeometric function \cite{Abr70}
in the Coulomb scattering wave function (\ref{cswf}).
After the integration over the spatial coordinates,
an integral remains that represents a hypergeometric function \cite{Abr70}.
This technique is similar to the evaluation of 
Bremsstrahlung matrix elements \cite{Nor54,Shy92}.
One obtains
\begin{eqnarray} 
\langle  \Psi^{(-)}_{\rm Coul}(\vec{k})  |  r^{-1} |
 \exp \left( i \vec{Q} \cdot \vec{r} \right)  \rangle
 & = & 
  \frac{4 \pi e^{-\frac{\pi}{2}\eta}}{(\vec{Q}-\vec{k})^{2}}
  \Gamma(1+i\eta) {}_{2}F_{1}(1,-i\eta;1; x)
\end{eqnarray}
with the argument
\begin{equation}
 x =  - 2\frac{\vec{k}\cdot(\vec{Q}-\vec{k})}{(\vec{Q}-\vec{k})^{2}}
\end{equation}
in the hypergeometric function ${}_{2}F_{1}$ which reduces to
the simple form
\begin{eqnarray}
 {}_{2}F_{1}(1,-i\eta;1;x) = (1-x)^{i\eta} 
\end{eqnarray}
for the given parameters. Combining the above results
the scattering amplitude assumes the form (\ref{fC0THres})
when the relation
\begin{equation}
   \Gamma(1+i\eta) = C_{0}(\eta) 
 \exp \left( \frac{\pi}{2}\eta +i\sigma_{0} \right)
\end{equation}
with the Coulomb phase $\sigma_{0}$ is used.

One application of this formula is the calculation of the
modified momentum amplitude (\ref{MMD}) with a Coulomb scattering wave function.
In this case one finds
\begin{eqnarray} 
 {\cal W}(\vec{q},\vec{k}) & = & 
 -  W(\vec{q}) \:  \frac{8 \pi \eta k}{Q^{2}-k^{2}}
  \frac{C_{0}(\eta)}{(\vec{Q}-\vec{k})^{2}}
 \exp \left\{ i\sigma_{0}(\eta) + i \eta \ln
  \left[\frac{Q^{2}-k^{2}}{(\vec{Q}-\vec{k})^{2}}\right] \right\}
\end{eqnarray}
where the indices $Bb$ of $\eta$ and $k$ have been suppressed for clarity.


\end{document}